\documentclass[12pt]{article}
\usepackage{epsf}
\newcommand{\vev}[1]{ \left\langle {#1} \right\rangle }
\newcommand{\gsim}{ \mathop{}_{\textstyle \sim}^{\textstyle >} }
\newcommand{\lsim}{ \mathop{}_{\textstyle \sim}^{\textstyle <} }

\usepackage{amssymb}
\usepackage{amsmath}
\usepackage{graphicx}
\usepackage{latexsym}
\usepackage{subfigure}

\begin{document}

\def\ds{\displaystyle}
\def\Slash#1{{\ooalign{\hfil$#1$\hfil\crcr\hfil$/$\hfil}}}

\newcommand{\rem}[1]{{\bf #1}}

\renewcommand{\theequation}{\thesection.\arabic{equation}}

\renewcommand{\thefootnote}{\fnsymbol{footnote}}
\begin{titlepage}

\def\thefootnote{\fnsymbol{footnote}}

\begin{center}


\hfill \today
\vskip .75in

{\Large \bf 
Muon g-2 and LHC phenomenology
in the $L_\mu-L_\tau$ gauge symmetric model }
\vskip .75in

{\large
Keisuke Harigaya$^1$, Takafumi Igari$^2$, Mihoko M. Nojiri$^{3,1}$, \\Michihisa Takeuchi$^4$, and Kazuhiro Tobe$^{2,5}$
}

\vskip 0.25in

{\em $^1$Kavli IPMU (WPI), TODIAS, University of Tokyo, Chiba, Kashiwa, 277-8583, Japan}

{\em $^2$Department of Physics, Nagoya University, Aichi, Nagoya 464-8602, Japan}

{\em $^3$Theory Center, KEK, Tsukuba, Ibaraki 305-0801, Japan}

{\em $^4$Theoretical Particle Physics and Cosmology Group, Department of Physics,
King's College London, London WC2R 2LS, UK}

{\em $^5$Kobayashi-Maskawa Institute for the Origin of Particles and
 the Universe, \\Nagoya University, Aichi, Nagoya 464-8602, Japan}

\end{center}
\vskip .5in

\begin{abstract}
In this paper, we consider phenomenology of a model with an $L_\mu-L_\tau$ gauge symmetry. 
Since the muon couples to the $L_\mu-L_\tau$
gauge boson (called $Z''$ boson), its contribution to the muon anomalous 
magnetic moment (muon g-2) can account for the discrepancy between the standard model prediction and 
the experimental measurements. On the other hand, the $Z''$ boson does not interact with the
electron and quarks, and hence there are no strong constraints from collider experiments
even if the $Z''$ boson mass is of the order of the electroweak scale.
We show an allowed region of a parameter space in the $L_\mu-L_\tau$ symmetric model, 
taking into account consistency with the electroweak precision measurements as well as 
the muon g-2. We study the Large Hadron Collider (LHC) phenomenology,
and show that the current and future data would probe the interesting parameter
space for this model.
\end{abstract}

KCL-PH-TH/2013-{\bf 37}

KEK-TH-1684

IPMU 13-0213

\end{titlepage}

\renewcommand{\thepage}{\arabic{page}}

\section{Introduction}
The Standard Model of elementary particles (SM) has been very successful 
in describing the nature at the electroweak (EW) scale.
Recently, the ATLAS and CMS collaborations at the Large Hadron Collider (LHC) 
have discovered a new particle~\cite{Aad:2012tfa,Chatrchyan:2012ufa},
which is consistent with the SM Higgs boson.  
This discovery also strengthens the correctness of the SM.
So far, no explicit evidence of physics beyond the 
SM has been reported from the LHC.

Several groups, however, have reported an anomaly of the muon anomalous magnetic moment 
$a_\mu=(g-2)/2$ (muon g-2), 
which has been precisely measured experimentally~\cite{Beringer:1900zz}
and compared with state-of-the-art theoretical predictions 
(for example, see \cite{Aoyama:2012wk, Czarnecki:2002nt, Hagiwara:2011af, 
Teubner:2010ah, Benayoun:2011mm, Jegerlehner:2011ti, Jegerlehner:2009ry, Davier:2010nc} and references therein).
The estimated discrepancies between the SM predictions and the measured value are consistently more than $3\sigma$, 
as listed in Table~\ref{tab:g-2}.

Although it is too early to conclude that this anomaly is evidence of new physics beyond the SM,
we expect new particles and interactions related with the muon sector
once we regard it as a hint of new physics.
Gauge interactions have been playing a central role to construct fundamental
models in particle physics history.
Following this line, in this paper, we purse the possibility that the
muon has a new gauge interaction beyond the SM%
\footnote{
If the new interaction is Yukawa-type one, new fermion or/and new scalar will be introduced. A well-known
example of this category is the minimal supersymmetric model~\cite{Moroi:1995yh}. Other models have been also
discussed in Refs.~\cite{Hambye:2006zn}.}.

\begin{table}[h]
\begin{center}
\begin{tabular}{|c|c|}
\hline
$a_{\mu}^{\rm Exp}~ [10^{-10}]$ & $\delta a_{\mu}=a_{\mu}^{\rm Exp}-a_{\mu}^{\rm SM}~[10^{-10}]$\\
\hline\hline
                	& $26.1\pm 8.0~(3.3\sigma)$~\cite{Hagiwara:2011af} \\
                        &  $31.6\pm 7.9~(4.0\sigma)$~\cite{Teubner:2010ah}\\
$11659208.9\pm 6.3$	& $33.5\pm 8.2~(4.1\sigma)$~\cite{Benayoun:2011mm} \\
	& $28.3\pm 8.7~(3.3\sigma)$~\cite{Jegerlehner:2011ti} \\
	& $29.0\pm 9.0~(3.2\sigma)$~\cite{Jegerlehner:2009ry} \\
	& $28.7\pm 8.0~(3.6\sigma)$~\cite{Davier:2010nc} \\
\hline
\end{tabular}
\end{center}
\caption{Measured muon g-2 ($a_\mu^{\rm Exp}$) and the estimated differences ($\delta a_\mu$) from 
the recent SM predictions in several references.}
\label{tab:g-2}
\end{table}

The discrepancy is of the same order as the contribution from the EW 
gauge bosons $W^{\pm}$ and $Z$, $a_{\mu}^{\rm EW}=(15.4\pm 0.2)\times 10^{-10}$~\cite{Czarnecki:2002nt}.
Assuming the anomaly is due to the quantum effects of the new particles, 
this discrepancy suggests that their masses should be at the EW scale, which is well within the reach of the LHC.
Thus, it is very interesting to study the phenomenology at the LHC.

The coupling of the new light gauge bosons to the electron and light
quarks are severely constrained by the
LEP~\cite{Abbiendi:2003dh,Abdallah:2005ph},
Tevatron~\cite{Aaltonen:2008ah,Abazov:2010ti} and
LHC~\cite{Aad:2012hf,Chatrchyan:2012oaa}.
Therefore, if the new gauge interaction is the flavor universal one such
as the $B-L$ gauge interaction,
the gauge coupling has to be small and hence the new gauge boson has to be very light as well 
in order to induce enough contributions to the muon g-2.
An explicit model of this category is the 
hidden photon model~\cite{Fayet:2007ua,Pospelov:2008zw}, and constraints on the 
hidden photon model have been studied in detail~\cite{Endo:2012hp,Davoudiasl:2012ig}.

Another possibility is a flavor-dependent gauge interaction.
If the gauge boson couples to the muon but
not to the electron nor quarks, the gauge interaction can explain muon g-2 
while keeping the consistency with the direct search results.
A simple candidate is based on anomaly free $L_\mu-L_\tau$ gauge symmetry 
\cite{He:1991qd, Baek:2001kca, Ma:2001md, Salvioni:2009jp, Heeck:2011wj},
where $L_{\mu}$ and $L_{\tau}$ are $\mu$ and $\tau$ lepton numbers, respectively. 
We consider this model in detail. 
In addition to the SM particles, this model has one extra gauge boson associated to 
the $L_\mu-L_\tau$ gauge symmetry, which we refer to as a $Z''$ boson.
Note that only the 2nd and the 3rd generation leptons couple to the $Z''$ gauge boson,
and hence the constraints from the direct search experiments are very weak.
\bigskip

Organization of this paper is the following. 
In the next section, we introduce a model with the $L_\mu-L_\tau$ gauge symmetry, 
which we refer to as a $Z''$ model.
In section 3, we study the parameter space of the $Z''$ model where the anomaly
of the muon g-2 can be explained. Since the mass of the $Z''$ gauge boson is expected 
to be of the EW scale if the gauge coupling is of order unity, the $Z''$ boson affects the 
EW precision observables. We investigate the effects and show
the parameter space consistent with the data.
In section 4, we study the LHC phenomenology.
We show that the $4\mu$ channel as well as the 
$2\mu 2\tau$ channel are effective for the $Z''$ boson search.
In section 5, we summarize our results.

\section{$Z''$ model}

The differences between two
lepton-flavor numbers $L_i-L_j~(i\neq j)$, where $L_{i}$ 
are lepton-flavor numbers, 
$L_i=(L_e,~L_\mu,~L_\tau)$, are anomaly free in the SM.
Therefore, the SM gauge 
symmetry ($G_{\rm SM}$) can be extended to $G_{\rm SM}\times U(1)_{L_i-L_j}$
without the addition of any exotic fermions, and such extensions are 
one of the minimal and economical $U(1)$ extensions of the SM.

In particular, the $L_\mu-L_\tau$ gauge symmetry is attractive because
it solves a problem of the muon g-2 without contradictions to other
experiments. The gauge boson, called $Z''$ boson, 
couples to the
2nd and 3rd generation leptons, so that it provides an extra contribution
to the muon g-2. Since the $Z''$ boson does not couple to the electron
nor any quarks, it avoids the strong constraints from the direct search
experiments.

In this paper, we consider a model based on 
$G_{\rm SM}\times U(1)_{L_\mu-L_\tau}$. 
The interactions of the $Z''$ boson are given by
\begin{equation}
\mathcal{L}_{int}=-g_{Z''}Z''_{\mu}\sum_{f=\mu,\tau,\nu_{\mu},\nu_{\tau}}Q''_f\bar{f}\gamma^{\mu}f,
\end{equation}
where $Q''_f$ is a $U(1)_{L_\mu-L_\tau}$ charge of a fermion 
$f$
as shown in Table \ref{charge}, and $g_{Z''}$ is the gauge coupling
constant of the $U(1)_{L_\mu-L_\tau}$ gauge symmetry.
We assume that the $L_{\mu}-L_{\tau}$ gauge symmetry is spontaneously broken 
and the $Z''$ boson becomes massive. The gauge coupling $g_{Z''}$ and the $Z''$ mass $m_{Z''}$
are the only free parameters in this model.

\begin{table}[h]
\begin{center}
\begin{tabular}{|c|c|c|c|c|c|c|c|}
\hline
particle & $L_2=(\nu_{\mu L},\mu_L)$ & $L_3=(\nu_{\tau L},\tau_L)$ &
$(\mu_R)^c$ & $(\tau_R)^c$ & $(\nu_{\mu R})^c$ & $(\nu_{\tau R})^c$ &others\\
\hline \hline
charge & +1 & $-1$ &$-1$ &+1 &$-1$ &+1 &0\\
\hline
\end{tabular}
\end{center}
\caption{Charges under the $L_\mu-L_\tau$ gauge symmetry. All fields are
in left-handed basis.}
\label{charge}
\end{table}

Right-handed neutrinos $\nu_{\ell R}\, (\ell=\mu,~\tau)$ can be light or heavy, 
depending on the model of neutrino masses in this framework. 
If they are light enough such that the decay mode $Z'' \to \nu_{\ell R}\bar{\nu}_{\ell R}$ is open, the branching ratio
$BR(Z''\rightarrow \mu^+\mu^-/\tau^+\tau^-)$ is about $1/4$, respectively. 
On the other hand, if they are heavy enough, $BR(Z''\rightarrow \mu^+\mu^- /\tau^+\tau^-)$ 
is about $1/3$, respectively.
Because of the smaller $\mu^+\mu^-/\tau^+\tau^-$
branching ratio,
it is more difficult to observe the $Z''$ boson signal
in the case that the right-handed neutrinos are light.
In order to be conservative, we assume that the right-handed neutrinos 
are light enough, and we refer to the model with this assumption as a $Z''$ model. 
We discuss a model with the correct neutrino masses and mixings in Appendix A.

In general, the $Z''$ gauge boson can mix with the $Z$ boson and photon 
since the $L_\mu-L_\tau$ symmetry is a $U(1)$ gauge symmetry.
Such mixings are naturally suppressed if the
$L_\mu-L_\tau$ gauge symmetry is embedded into 
a non-Abelian symmetry at the more fundamental
level~\cite{Salvioni:2009jp}.
Therefore, in this paper, we assume that the $U(1)$ mixing effect is negligible.

\section{Muon g-2 and electroweak precision observables}

\begin{figure}[t]
\centerline{
\includegraphics[width=7cm]{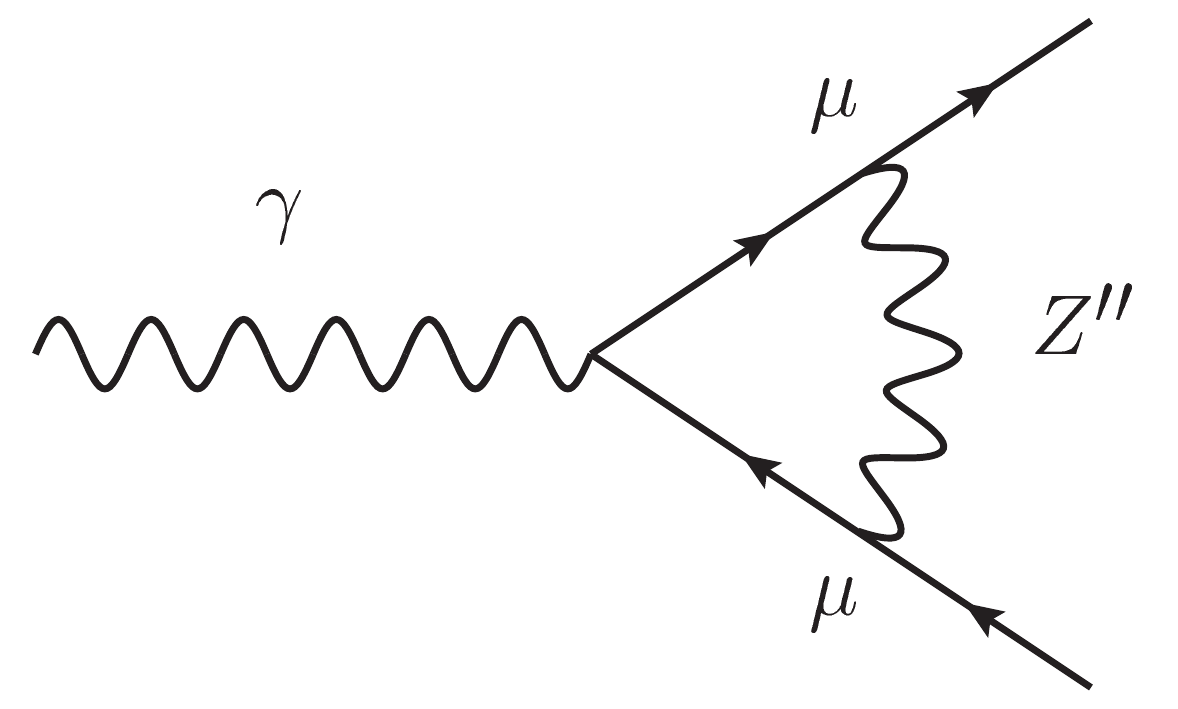}
}
\caption{Feynman diagram for muon g-2, mediated by the $Z''$ gauge boson.
}
\label{muong2}
\end{figure}

\begin{figure}[b]
\centerline{
\includegraphics[width=8cm]{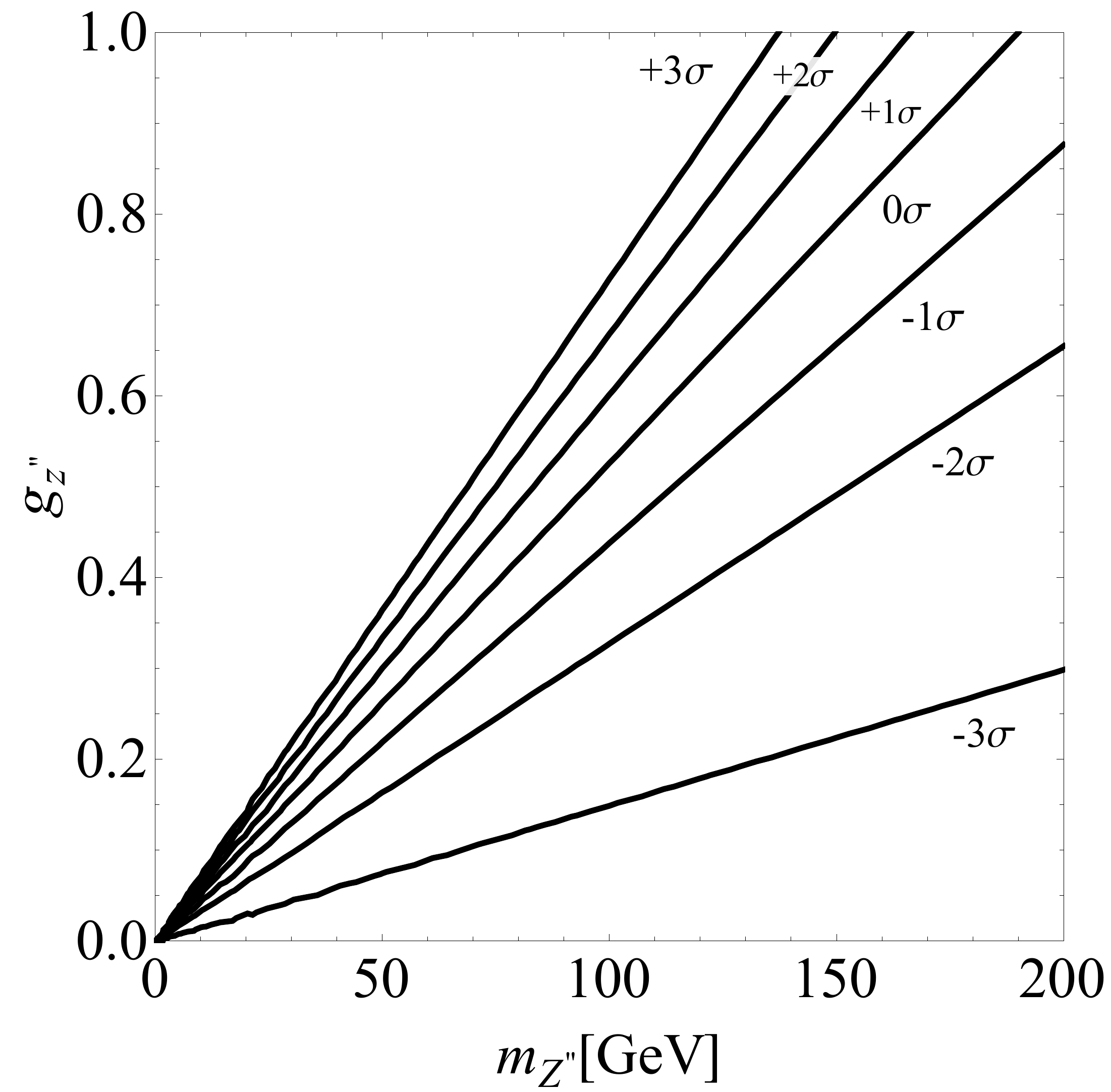}
}
\caption{
Contours of the standard deviations for muon g-2 with
the $Z''$ contribution ($\delta a_\mu$) in $(m_{Z''},g_{Z''})$ plane.
}
\label{amu}
\end{figure}

In this section, we study the parameter space of the $Z''$ model 
in which the measured muon g-2 is explained while satisfying the
constraints from the EW precision observables.
As shown in Table~\ref{charge}, the $Z''$ gauge boson interacts 
with the muon through the $L_\mu-L_\tau$ gauge interaction.
The new contribution to the muon g-2 ($\delta a_\mu$) is induced through 
a Feynman diagram depicted in Figure~\ref{muong2}, and it is given by the following expression~\cite{Ma:2001md},
\begin{eqnarray}
\delta a_\mu = \frac{g_{Z''}^2}{12\pi^2} \frac{m_\mu^2}{m_{Z''}^2}
\simeq 2\times 10^{-9}~\left(\frac{g_{Z''}}{0.5}\right)^2
\left(\frac{100~{\rm GeV}}{m_{Z''}}\right)^2.
\end{eqnarray}
Here, we have assumed that $m_{Z''}\gg m_\mu$.
Figure~\ref{amu} shows the dependence of $\delta a_\mu$ on the two model parameters.
It can be seen that the effect of the $Z''$ boson with $g_{Z''}=O(1)$ and $m_{Z''}=O(100)$ GeV  
compensates the $3\sigma$ deviation observed in the muon g-2 measurement. 


Since $Z''$ couples to the 2nd and 3rd generation leptons ($l, \nu_l$), it induces extra contributions 
to $Z\bar{l}l$ and $W^+l^-\bar{\nu}_l$ vertices. 
The $Z''$ boson effects appear on the EW precision observables through these vertex corrections.
The effective vertex $Z_\mu \bar{f} f $ is given by
\begin{eqnarray}
\frac{ig}{c_W}\gamma_\mu(g^f_L P_L +g^f_R P_R),
\end{eqnarray}
where 
\begin{eqnarray}
g^f_L =  (T^3_f-Q_f s_W^2)(1+\Delta),\quad \quad
g^f_R = -Q_f s_W^2 (1+\Delta).
\end{eqnarray}
Note that one-loop corrections are parametrized
by $\Delta$, which is independent of $f$ ($f=\mu, \tau, \nu_{\mu}, \nu_\tau$) if the lepton masses are neglected.
$\Delta$ is given by~\cite{Carone:1994aa}
\begin{eqnarray}
\Delta&=&\Delta^{(1)}+\delta Z,\\
\Delta^{(1)}&=&-\frac{g_{Z''}^2}{8\pi^2}{\rm Re}\left[
q^2 \left\{ C_{0}+C_{11}+C_{23}-C_{22} \right\}
-2(1-\epsilon)^2 C_{24} \right](Z'',\mu,\mu;p,q),\\
\delta Z &=&-\frac{g_{Z''}^2}{8\pi^2}(1-\epsilon)(B_0+B_1)(Z'',\mu;p^2=m_\mu^2),
\end{eqnarray}
where $\Delta^{(1)}$ is an one-loop vertex correction and $\delta Z$ is a 
counter term contribution from the wave function renormalization of the
leptons.
$p$ and $q$ are external momenta of the muon and the $Z$ boson,
respectively.
$C_{XX}$ and $B_X$, so-called the
Passarino-Veltman functions, are given in Appendix B.
The explicit form of $\Delta$ is
\begin{eqnarray}
\Delta(q^2)&=&-\frac{g_{Z''}^2}{8\pi^2}\left[
\frac{7}{4}+\delta +\left(\delta+\frac{3}{2}\right)\log\delta \right. \nonumber\\
&&
\left.\left.
+(1+\delta)^2\left\{\rm{Li}_2\left(\frac{\delta}{1+\delta}\right)
+\frac{1}{2}\log^2\left(\frac{\delta}{1+\delta}\right)
-\frac{\pi^2}{6}\right\} \right] \right.,
\end{eqnarray}
where $\delta\equiv\frac{m_{Z''}^2}{q^2}$ and $\rm{Li}_2\left(x\right)\equiv-\int_0^x dt \frac{\log(1-t)}{t}$ is 
the Spence function.
Here, we have neglected the muon mass.

\begin{figure}
\centerline{
\includegraphics[width=8cm]{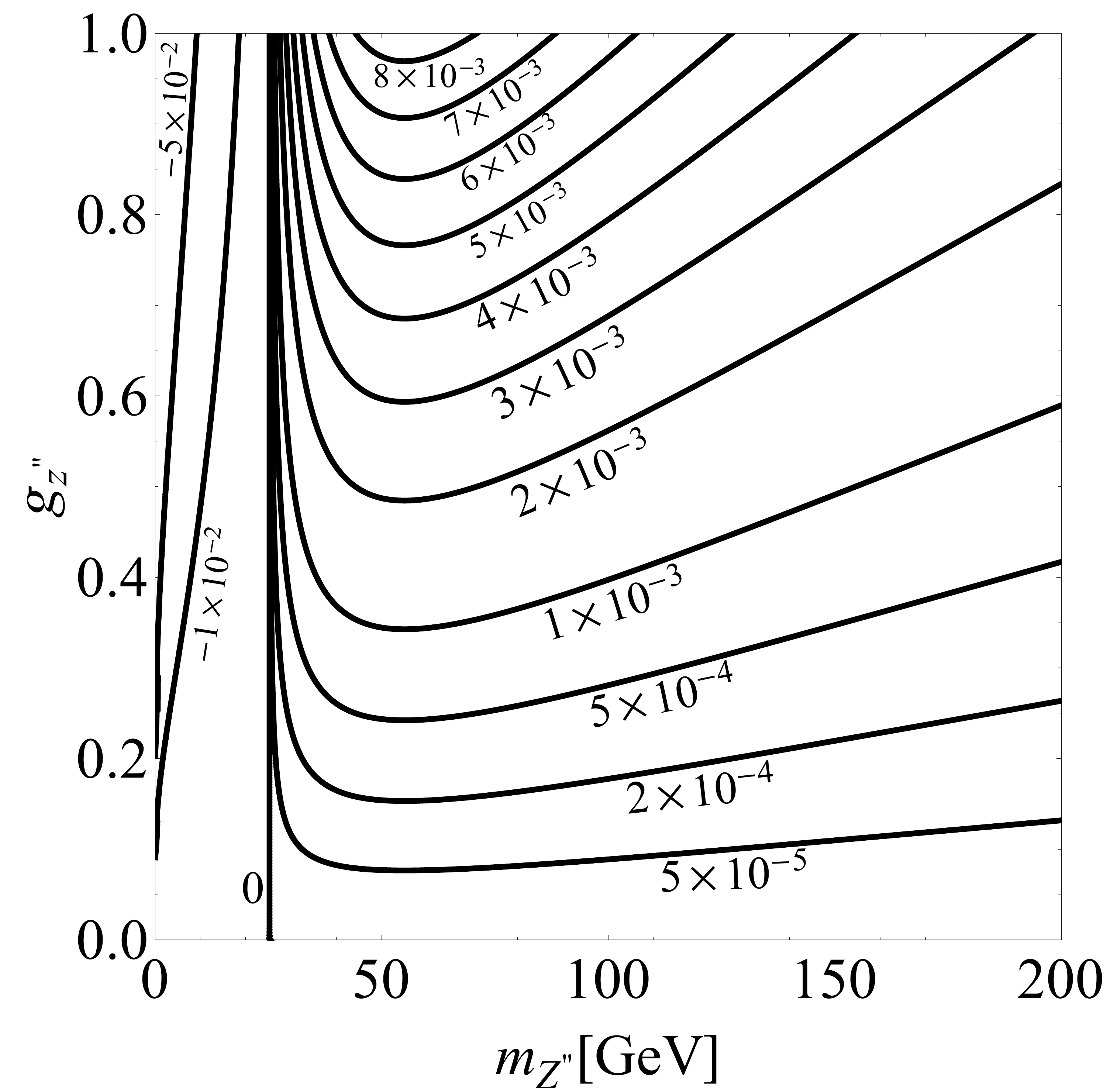}
}
\caption{The vertex correction $\Delta$ at the $Z$-pole ($q^2=m_Z^2$) is shown as a 
function of $m_{Z''}$ and $g_{Z''}$.}
\label{Delta}
\end{figure}

Figure~\ref{Delta} shows the numerical values of the vertex correction
 at the $Z$-pole $\Delta(q^2=m_Z^2)$ as a function of $m_{Z''}$ and $g_{Z''}$. 
For example, with $g_{Z''}=0.3$ and  $m_{Z''}=60$ GeV ($m_{Z''}=80$ GeV),
we obtain $\Delta(m_Z^2)=7.6\times 10^{-4}~(6.7\times10^{-4})$.

Similarly, we can calculate an one loop correction to $W^+ l^- \bar{\nu}_l$ vertex ($l=\mu,~\tau$)
via the $Z''$ gauge boson, and the effective vertex is given by
\begin{eqnarray}
\frac{ig}{\sqrt{2}}\gamma_\mu k_L^l P_L,
\end{eqnarray}
where
\begin{eqnarray}
k_L^l=1+\Delta_W.
\end{eqnarray}
We obtain $\Delta_W=\Delta$ by neglecting the $\mu$ ($\tau$) and $\nu_\mu$ ($\nu_\tau$) masses.

One may be worried that
the $Z''$ boson contribution to $W\mu\nu_\mu$ 
vertex may affect the muon decay, $\mu\rightarrow e\nu_\mu \bar{\nu}_e$.  
However, the momentum transfer of the virtual $W$ boson is negligible in the muon decay 
compared to the mass scale of the $Z''$ boson.
Therefore, the effect of the $Z''$ boson decouples in the muon decay.

We calculate the EW precision observables listed in Table~\ref{EWObs}. We adopt formulas in, 
for example, Refs.~\cite{Cho:2011rk,Cho:1999km,Hagiwara:1994pw} for the calculation.
We perform a $\chi^2$ fit by varying input parameters $m_t,~m_h,~\Delta \alpha_{\rm had}^{(5)}$ and $\alpha_s$,
in order to identify the parameter region consistent with the data. 
The $\chi^2$ is defined by

\begin{equation}
\chi^2=\sum_{i,~ j}\left(\frac{O_i^{\rm Exp}-O_i^{\rm Model}}{\sigma^{\rm Exp}_i}
\right)^2
(\rho^{-1})_{ij}
\left(\frac{O_j^{\rm Exp}-O_j^{\rm Model}}{\sigma^{\rm Exp}_j}
\right)^2,
\end{equation}
where $O^{\rm Exp}$, $\sigma^{\rm Exp}$ and $\rho$ are the measured
value, the $1~\sigma$ error and the correlation coefficient matrix
of the observables, respectively, taken from Refs.~\cite{Beringer:1900zz,Hagiwara:2011af,ALEPH:2005ab}, and
$O^{\rm Model}$ is the theoretical prediction.
In Table~\ref{EWObs}, we show the result at the best fit point for the SM
and the results at the sample points with $m_{Z''}=(60,~80)$ GeV and $g_{Z''}=0.3$
for the $Z''$ model.

\begin{table}[p]
\begin{center}
\begin{tabular}{|c|c||c|c||c|c|c|c|}
\hline
 & data & SM fit & pull & $Z''$ model& pull & $Z''$ model & pull\\
\hline
$\Gamma_Z$(GeV)& 2.4952(23) & 2.4953& -0.06 & 2.4961& -0.4 &2.4960 & -0.3\\
$\sigma_h^0$ (nb) & 41.541(37) &41.480 & 1.7&41.454  & 2.3 &41.457 & 2.3\\
$R_e$ & 20.804(50) & 20.739& 1.3 & 20.739& 1.3 & 20.739 & 1.3\\
$R_\mu$ & 20.785(33) & 20.739& 1.4 & 20.708& 2.3 & 20.712 &2.2\\
$R_\tau$ & 20.764(45) & 20.787& -0.5 & 20.755& 0.2 & 20.759 &0.1\\
$A_{\rm FB}^{0,e}$ & 0.0145(25) & 0.0162& -0.7 & 0.0162& -0.7 & 0.0162 &-0.7\\
$A_{\rm FB}^{0,\mu}$ & 0.0169(13) & 0.0162& 0.5 & 0.0162 & 0.5& 0.0162 &0.5\\
$A_{\rm FB}^{0,\tau}$ & 0.0188(17) & 0.0162& 1.5 & 0.0162& 1.5 &0.0162 &1.5\\
{\small $\tau$ pol.:} & & & & & & &\\
$A_\tau$ & 0.1439(43) & 0.1472& -0.8 &0.1472 & -0.8 &0.1472 & -0.8\\
$A_e$ & 0.1498(49) & 0.1472& 0.5 & 0.1472& 0.5 & 0.1472 & 0.5\\
{\small $b$, $c$ quarks:} & & & & & & &\\
$R_b$ & 0.21629(66) & 0.21579& 0.8 &0.21579  & 0.8 &0.21578 &0.8\\
$R_c$ & 0.1721(30)& 0.1722& -0.05 & 0.1722& -0.05 &0.1722 & -0.05\\
$A_{\rm FB}^{0,b}$ & 0.0992(16) &0.1032 &-2.5 &0.1032 &-2.5 & 0.1032 &-2.5\\
$A_{\rm FB}^{0,c}$ & 0.0707(35) & 0.0737 & -0.9 & 0.0737& -0.9 & 0.0737 & -0.9\\
$A_b$ & 0.923(20) & 0.935  & -0.6 & 0.935& -0.6 & 0.935 & -0.6\\
$A_c$ & 0.670(27) &0.668 & 0.08 & 0.668& 0.08 &0.668 & 0.08\\
{\small SLD:} & & & & & & &\\
$A_e$ & 0.1516(21) &0.1472 &2.1  &0.1472 &2.1 &0.1472 & 2.1\\
$A_\mu$ & 0.142(15) &0.1472 &-0.3 &0.1472 &-0.3 &0.1472 & -0.3\\
$A_\tau$ & 0.136(15) &0.1472 &-0.7 &0.1472 &-0.7 &0.1472 &-0.7\\
{\small W boson:} & & & & & & &\\
$M_W$ (GeV) & 80.385(15)
& 80.362& 1.5 & 80.362& 1.5 &80.362 &1.5\\
$\Gamma_W$ (GeV) & 2.085(42)& 2.091& -0.1 & 2.091&-0.2 &2.091 & -0.2\\
muon g-2: & & & & & & &\\
$\delta a_\mu(10^{-9})$ & 2.61(0.80) & 0 & 3.3 &2.36 &1.1 &1.33 &1.1\\
\hline
Inputs & & & & & & &\\
$\Delta \alpha_{\rm had}^{(5)}(M_Z^2)$ & 0.02763(14) & 0.02760&0.2 &0.02760 &0.2&0.027560 &0.2\\
$\alpha_s(M_Z)$ & 0.1184(7) & 0.1184&0.0 & 0.1184& 0.0 &0.1184& 0.0\\
$m_t$ (GeV) & 173.1(0.9)
&173.7 &-0.6 &173.7 &-0.6&173.7 &-0.6\\
$m_h$ (GeV) &  125.9 (0.4)& 125.9& 0& 125.9& 0&125.9 &0\\
\hline
$m_{Z''}$ (GeV) &- &- &- & 60 & -& 80 &-\\
$g_{Z''}$ &- &- &- & 0.3 &- & 0.3 &-\\
\hline
$\chi^2/(d.o.f) $ & & 35.1/(22)& &29.2/(22)& & 31.0/(22)&
\\
\hline
\end{tabular}
\end{center}
\caption{The EW precision data and theoretical predictions of EW precision observables.
The experimental data are taken from Ref.~\cite{ALEPH:2005ab} except that $M_W$, $\Gamma_W$, $m_t$
and $m_h$ are from Ref.~\cite{Beringer:1900zz}, and $\delta a_\mu$ and $\Delta \alpha_{\rm had}^{(5)}$
are from Ref.~\cite{Hagiwara:2011af}. The best fit values of the SM and 
sample points for $Z''$ model, $m_{Z''}=(60,~80)$ GeV and $g_{Z''}=0.3$ are shown.}
\label{EWObs}
\end{table}

The $Z''$ effects increase the partial $Z$ decay widths
of $l\bar{l}$ mode $\Gamma_{ll}$ ($l=\mu, \tau, \nu_\mu$ and $\nu_\tau$) in the interesting 
parameter region while the $Z$ total decay width $\Gamma_Z$ does not increase significantly 
since the hadronic contributions $\Gamma_{\rm had}$ are dominant. 
Similarly, the effect on the $W$ total decay width is also negligible.
On the other hand, the effects on 
$\sigma_h^0=\frac{12\pi}{m_Z^2}\frac{\Gamma_{ee} \Gamma_{\rm had}}{\Gamma_Z^2}$ 
and $R_\mu=\frac{\Gamma_{\rm had}}{\Gamma_{\mu\mu}}$
can be significant. As shown in Table~\ref{EWObs}, the observed value of $\sigma_h^0$ tends to be
larger than the SM value. Adding $Z''$ contribution results in worse fittings.
Similarly, since the observed value of $R_\mu$ is larger than the SM fitted value, adding the $Z''$ contribution
makes the fit further worse.\footnote{
We also note that the $Z''$ contribution does not affect the left-right asymmetry $A_{\mu,\tau}$,
because the $Z''$ universally contributes to the left- and right-handed $\mu$ and $\tau$.}
As a result, large vertex corrections from the $Z''$ boson are disfavored.

\begin{figure}[h]
\centerline{
\includegraphics[width=8cm]{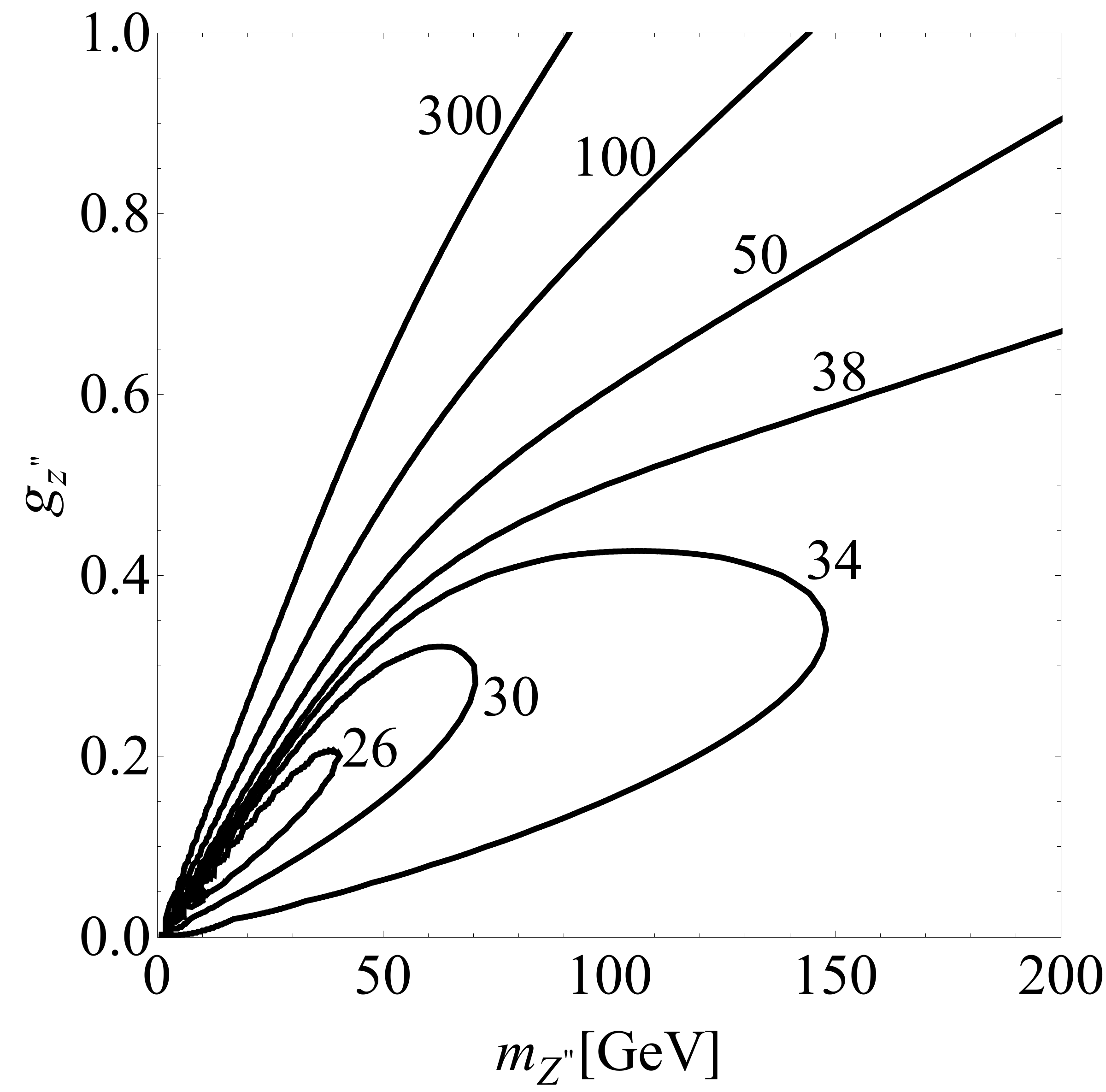}
}
\caption{The total $\chi^2$ in the $(m_{Z''}, g_{Z''})$ plane.
}
\label{chi2_EW}
\end{figure}

In Figure~\ref{chi2_EW}, we show $\chi^2$ of the $Z''$ model as a function of
$m_{Z''}$ and $g_{Z''}$. As can be seen from the figure, small gauge coupling $g_{Z''}<0.4$ and relatively 
light $Z''$ boson $m_{Z''}<100$ GeV are favored.

\newpage

\section{LHC phenomenology}
In this section, we study the phenomenology of the $Z''$ model at the LHC and investigate 
whether the current and future LHC results can constrain or 
discover the $Z''$ boson in the region which is favored by the EW precision 
measurement as well as the muon g-2 shown in the previous section.

\begin{figure}[h]
\centerline{
\includegraphics[width=10cm]{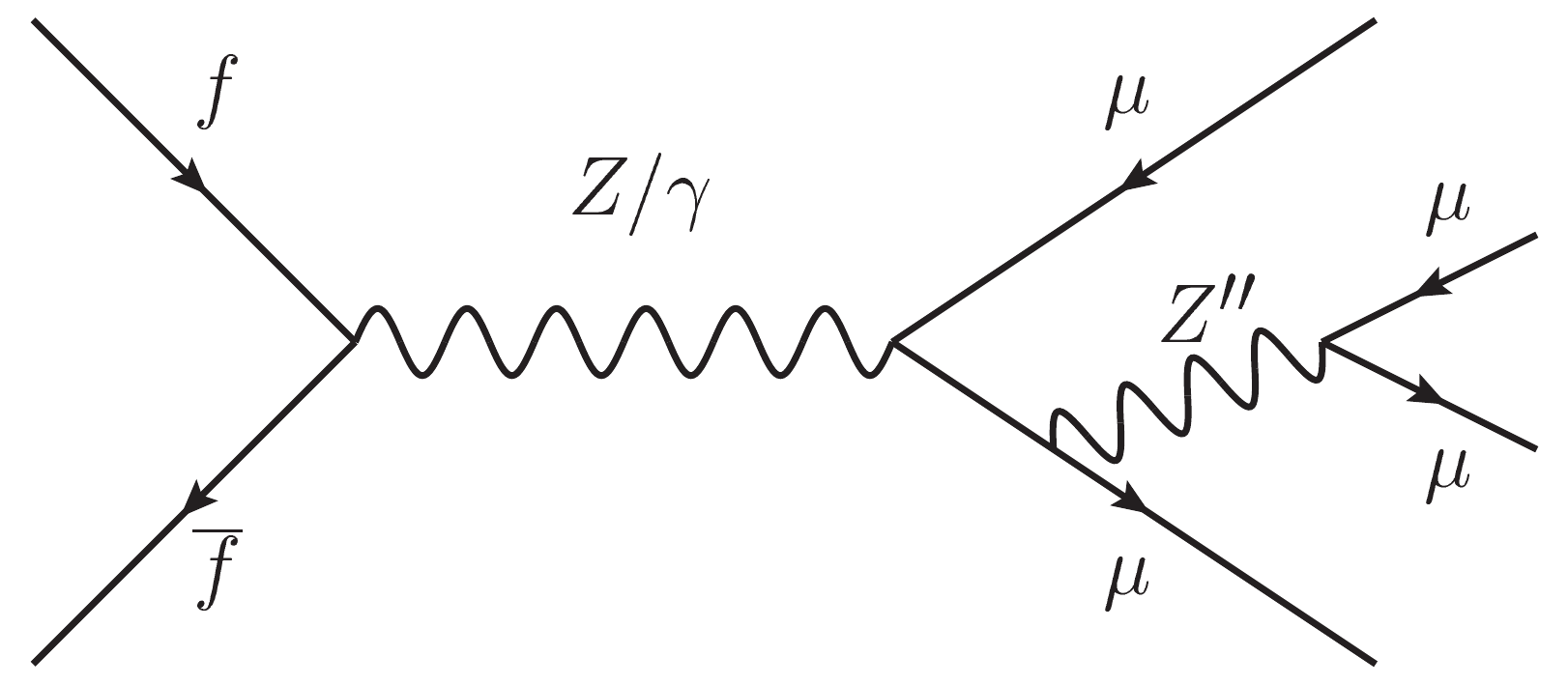}
}
\caption{Feynman diagram for a typical $Z''$ boson production process at the tree-level. 
}
\label{ZpFeyn}
\end{figure}

Relatively light $Z''$ bosons can be produced at $e^+e^-$, $p\bar{p}$ and $pp$ collisions.
The event including the decay is typically described by the diagram depicted in Figure~\ref{ZpFeyn}. 

Since the $Z''$ boson only couples to $\mu$, $\tau$, $\nu_\mu$ and $\nu_\tau$, 
its effects only appear in the specific final states. Table~\ref{Zp_patern} lists 
the final states where the $Z''$ boson contributes.
In particular, the 4 lepton modes involving $e^\pm$ are not affected.

\begin{table}[h]
\begin{center}
\begin{tabular}{l|c}
\hline
final state & $Z''$ effects\\
\hline \hline
$4\mu$, $4\tau$, $2\mu 2\tau$, $2\mu+E_{T,\rm{miss}}$, $2\tau+E_{T,\rm{miss}}$ & yes\\
$4e$, $2e2\mu$, $2e 2\tau$, $2e+E_{T,\rm{miss}}$ & no \\
\hline
\end{tabular}
\end{center}
\caption{List of 4 lepton final state processes relevant to the $Z''$ boson effects. 
Missing transverse energy, $E_{T,\rm{miss}}$, is originated from neutrinos.}
\label{Zp_patern}
\end{table}

Table~\ref{cross} lists the cross sections of typical processes for the SM and $Z''$ model for 
$m_{Z''}=80$ GeV. The gauge coupling is fixed as $g_{Z''}=0.3$ 
throughout this section unless otherwise stated.
We see that there are no constraint we can set  on the $Z''$ model 
from the LEP~\footnote{In Ref.~\cite{Ma:2001md}, authors discussed the LEP bound, however, we
found their limit is too optimistic.} 
 nor from the Tevatron due to the small cross sections.
On the other hand, 4 $\mu$ final states have been already observed 
at 7-8~TeV LHC with $\int dt L=25~\rm{fb}^{-1}$,
and a few fb difference in the cross section would be or become measurable~\cite{Ma:2001md}.
For the $2\mu2\tau$ mode, it is difficult to constrain the $Z''$ model 
from the current data due to low $\tau$ identification efficiency though 
it would be possible with $\int dt L=300-3000~{\rm fb}^{-1}$ at $\sqrt{s}=14$ TeV.

In this section the results based on four signal samples with $m_{Z''}=60$, 80, 90, and 100~GeV are shown.
Note that the generated signals  also include the SM  and the interference contributions.
We perform a parton level calculation using Calchep-3.4~\cite{Belyaev:2012qa}
and interface the events to Pythia-6.4.25~\cite{Sjostrand:2006za}.
The detector effect is simulated with Delphes-2.0.5~\cite{Ovyn:2009tx}.

\begin{table}[h]
\begin{center}
\begin{tabular}{ll|c|c}
\hline
\multicolumn{2}{c|}{process} & 
\multicolumn{2}{c}{cross section [fb]}\\
\cline{3-4}
\multicolumn{2}{c|}{} 
& SM & $Z''$ model ($m_{Z''}=80$~GeV) \\
\hline\hline
LEP ($\sqrt{s}=200$~GeV) & 
$e^+e^-\rightarrow 4\mu$ & 3.8 & 3.8  \\
\hline
Tevatron ($\sqrt{s}=1.96$~TeV) & 
$p\bar{p}\rightarrow 4\mu$ & 3.4 & 3.6	\\
\hline
LHC ($\sqrt{s}=8$~TeV) & 
$pp \rightarrow 4\mu$ & 14 & 15 \\
& $pp\rightarrow 2\mu2\tau$ & 29 & 30 \\
\hline
LHC ($\sqrt{s}=14$~TeV) &
$pp \rightarrow 4\mu$ & 27 & 28 \\
& $pp\rightarrow 2\mu2\tau$ & 57 & 59 \\
\hline
\end{tabular}
\end{center}
\caption{Cross sections in typical processes where the $Z''$ boson contributes,
where $p_{T, l}>5$ GeV and $m_{l^-l^+}>5$ GeV ($l=\mu~{\rm and}~\tau$) are required. 
The numbers for the $Z''$ model are for $m_{Z''}= 80$ GeV and $g_{Z''}=0.3$.
}
\label{cross}
\end{table}

\subsection{4 lepton channels at $\sqrt{s}=7-8$ TeV}
\label{subsec:4mu_8TeV}

Both CMS~\cite{CMS:2012bw} and ATLAS~\cite{ATLAS_Z4l} collaborations have reported 
the measurements of $Z$ decays to four leptons at $\sqrt{s}= 7$ and 8 TeV. 
Their measurements would be sensitive to the light $Z''$ boson.
First, we consider 
how strongly  the existence of  the $Z''$ boson is  constrained by the ATLAS data~\footnote{
The CMS has similar analysis in Ref.~\cite{CMS:2012bw}
and their result, however, is based on data collected at
$\sqrt{s}=7$ TeV. On the other hand, the ATLAS result is based on much larger set of data with integrated luminosities of 4.6 fb$^{-1}$
at $\sqrt{s}=7$ TeV and 20.7 fb$^{-1}$ at $\sqrt{s}=8$ TeV. Therefore, we concentrate on the ATLAS analysis.
}.
In the ATLAS analysis~\cite{ATLAS_Z4l}, they search for the production of four leptons: $e^+e^-e^+e^-~(4e)$,
$\mu^+\mu^-\mu^+\mu^-~(4\mu)$ and $e^+e^-\mu^+\mu^-~(2e2\mu)$ at the $Z$ resonance.
We summarize the set of selection cuts they have used as follows:
\begin{enumerate}
\item four isolated leptons, which have two opposite sign and same-flavor di-lepton pairs, where
$p_{T,\mu}>4$ GeV and $|\eta_\mu|<2.7$ ($p_{T,e}>7$ GeV and $|\eta_e|<2.47$).
\item the leading three leptons must have $p_{T,\ell}>20,~15$, and 8 GeV, and if the third ($p_T$-ordered) lepton
is an electron it must have $p_{T,e_3}>10$ GeV.
\item  the four leptons are required to be separated as $\Delta R_{\ell \ell}>0.1$.
\item  the invariant masses of the same-flavor and opposite-sign leptons are required to have $m_{l^+l^-}>5$ GeV.
\item  $m_{12}>20$ GeV and $m_{34}>5$ GeV, 
where $m_{12}$ is the invariant mass of the same flavor and opposite sign di-lepton pair 
which is the closest to the $Z$ boson mass among the possible combinations, 
while the other one is called $m_{34}$.
\item the invariant mass of the four leptons is in the $m_Z$ window, 80 GeV $<m_{4l}< 100$ GeV.
\end{enumerate}

\begin{figure}[b]
\centerline{
\subfigure[]{\includegraphics*[width=8cm]{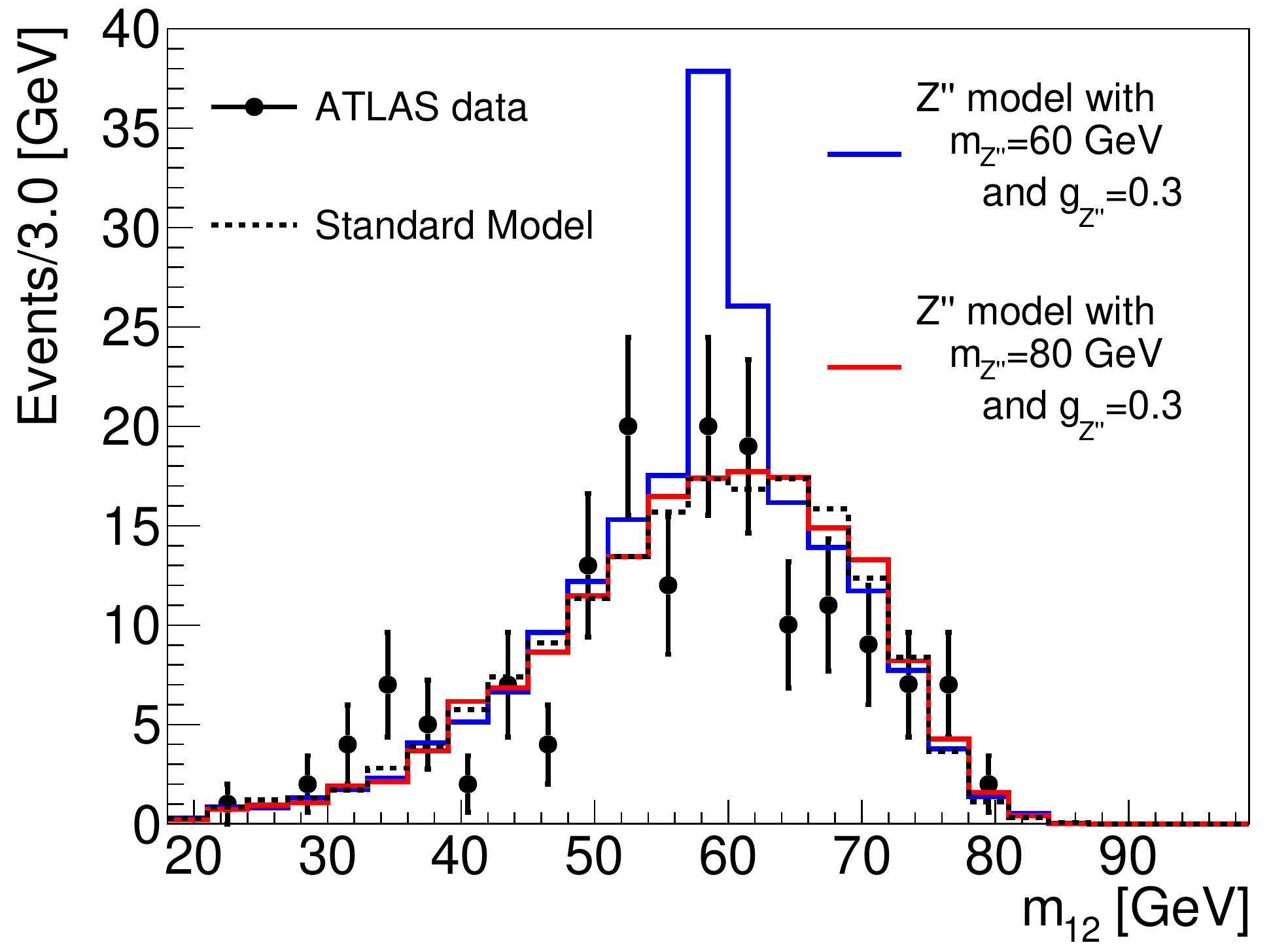}
}
\subfigure[]{\includegraphics*[width=8cm]{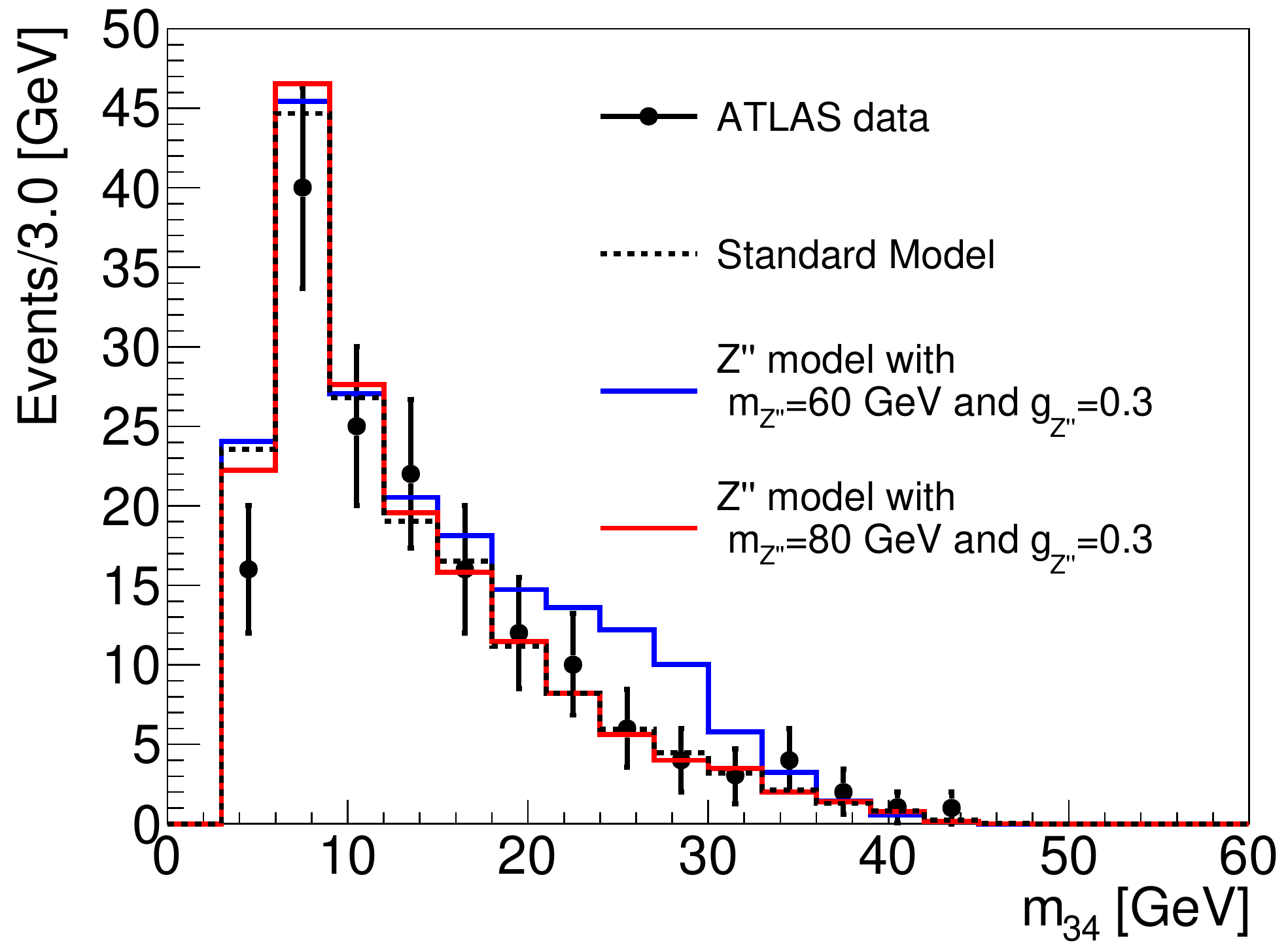}
}
}
\caption{The $m_{12}$ and $m_{34}$ distributions
for the SM (dashed) and for the $Z''$ models with $m_{Z''}=60$ GeV (blue) and $80$ GeV (red).
All channels ($4e$, $2e2\mu$ and $4\mu$) are summed up.
Combined results for the integrated luminosities of 4.6 fb$^{-1}$ at $\sqrt{s}=7$ TeV 
and 20.7 fb$^{-1}$ at $\sqrt{s}=8$ TeV are shown.
}
\label{fig:m12m34}
\end{figure}

In the following we compare our simulation results with the ATLAS results.
In order to adjust $K$-factor, acceptance and efficiency factors in the simulation, 
we introduce a constant normalization factor in each of the channels ($4e$, $2e2\mu$ and $4\mu$) 
to match our LO SM results and the expected numbers of events in Table 4 in Ref.~\cite{ATLAS_Z4l}, which is obtained by NLO Monte Carlo program POWHEG\cite{POWHEG} and data driven acceptance estimations.
We use the same factors for the $Z''$ models.

In Figure~\ref{fig:m12m34} we show the di-lepton invariant mass $m_{12}$ distributions  
and $m_{34}$ distributions in the SM and $Z''$ models with $m_{Z''}=60$~GeV (left panel) and $80$~GeV (right panel).
In this section, combined results for integrated luminosities of 4.6 fb$^{-1}$ at $\sqrt{s}=7$ TeV and for 20.7 fb$^{-1}$ at 
$\sqrt{s}=8$ TeV are shown.
All channels ($4e$, $2e2\mu$ and $4\mu$) are summed up for these plots
so that we can directly compare them with  Figure~3(e) and 3(f) in Ref.~\cite{ATLAS_Z4l}.

For the $Z''$ model with $m_{Z''}=60$ GeV, 
a large excess should be seen around $m_{12}\simeq m_{Z''}$ in the $m_{12}$ distribution.
It is from the on-shell decay $Z \to Z'' \ell^+ \ell^-$ followed by $Z'' \to \ell^+ \ell^-$.
We also see a small excess around $m_{34}=20-30$ GeV in the $m_{34}$ distribution, and it corresponds to
$m_Z - m_{Z''}$. On the other hand, we don't see significant deviations for $m_{Z''}=80$ GeV.
Table~\ref{table_m12} shows the expected numbers of events in several $m_{12}$ and $m_{34}$ ranges
for the $Z''$ model with $m_{Z''}=60$ GeV  ($N_{Z'', 60}$) and $80$~GeV ($N_{Z'', 80}$) , and 
for the SM ($N_{\rm SM}$).  The ``significance'' value $\sigma_{Z''}$, which is defined by 
$(N_{Z''}-N_{\rm SM})/\sqrt{N_{\rm SM}}$ and represents the deviation from the SM, is also shown.
For $m_{Z''}=60$ GeV, $\sigma_{Z'',60}$ values are about 5.1 and 4.1 in the range of $m_{12}=57-63$ GeV 
and of $m_{34}=18-33$ GeV, respectively, and shows the clear deviation from the SM prediction,
while $\sigma_{Z'',80}$ is smaller than 1 and not statistically significant.
\bigskip

\begin{table}[t]
\begin{center}
\begin{tabular}{cc||c|rr|rr}
\hline
&& $N_{\rm SM}$ & $N_{Z'',60}$ & $\sigma_{Z'',60}$& $N_{Z'',80}$ &$\sigma_{Z'',80}$ \\
\hline
&$(51, 57)$~GeV & 29.1 & 32.8 & 0.7 & 29.9 & 0.1\\
&$(57,63)$~GeV & 34.2 & 63.9 & 5.1 & 35.1 & 0.2\\
$m_{12}$ & $(63,69)$~GeV & 33.2 & 30.1 & -0.5 & 32.3 &-0.2\\
&$(69,75)$~GeV & 20.7 & 19.5 & -0.3 & 21.5 & 0.2 \\
&$(75,81)$~GeV & 4.7  & 5.2 & 0.2 & 5.9  & 0.5 \\
\hline
&(3,18)~GeV& 130.6	& 135.2	& 0.4 & 131.8	& 0.1 \\ 
$m_{34}$ &(18,33)~GeV & 33.0	& 56.4	& 4.1 & 32.8	& -0.0 \\
&(33,48)~GeV& 4.5	& 5.4	& 0.4 & 4.3	& -0.1 \\
\hline
\end{tabular}
\end{center}
\caption{Event numbers in several $m_{12}$ and $m_{34}$ ranges. The luminosities of 4.6 fb$^{-1}$ at $\sqrt{s}=7$ TeV and 20.7 fb$^{-1}$ at $\sqrt{s}=8$ TeV are
combined and event numbers in all channels ($4e$, $2e2\mu$ and $4\mu$) are summed up, as studied in Ref.~\cite{ATLAS_Z4l}.
$N_{\rm SM}$ and $N_{Z'',60}$ ($N_{Z'',80}$) are numbers of events in the SM and the $Z''$ model with
 $m_{Z''}=60$ GeV (80 GeV), respectively.
 We also show the significance $\sigma_{Z''}=(N_{Z''}-N_{\rm SM})/\sqrt{N_{\rm SM}}$.
}
\label{table_m12}
\end{table}

\begin{table}[h]
\begin{center}
\begin{tabular}{l||r|r|r}
\hline
 &SM & $Z''$ model ($m_{Z''}=60$~GeV) & $Z''$ model  ($m_{Z''}=80$~GeV)\\
\hline\hline
$\chi^2$/(d.o.f) in $m_{12}$& 33.1/(19) & 47.1/(19) & 34.1/(19) \\
$\chi^2$/(d.o.f) in $m_{34}$& 6.9/(14) & 26.6/(14) & 6.5/(14)\\
\hline
\end{tabular}
\end{center}
\caption{$\chi^2$ in the $m_{12}$ and $m_{34}$ distributions 
in the SM and the $Z''$ models with $m_{Z''}=60$ and 80~GeV.
}
\label{chi2}
\end{table}

We also compute the $\chi^2$ values defined by
\begin{eqnarray}
\chi^2=\sum_{i}\left(\frac{N_{th}^i-N_{\rm DATA}^i}{\sigma_i}\right)^2,
\end{eqnarray}
where $N_{\rm th}^i$ is the expected number of events in the i-th bin
for the theoretical models (the SM and the $Z''$ models),
$N_{\rm DATA}^i$ and $\sigma_i$ are the number of events observed in the i-th bin for the data
and the corresponding statistical error, respectively. The data are obtained from 
the ATLAS analysis~\cite{ATLAS_Z4l}.

Table~\ref{chi2} shows the $\chi^2$ values for 
the $m_{12}$ and $m_{34}$ distributions for the
SM and the $Z''$ models.
We use 19 bins for the $m_{12}$ distribution and 
14 bins for the $m_{34}$ distribution to calculate 
the $\chi^2$.
Thus, the degree of freedom (d.o.f) of the $\chi^2$ is 19 (14) for the $m_{12}$ ($m_{34}$) distribution. 
We see that the $\chi^2$ for the $Z''$ model with $m_{Z''}=60$ GeV 
is much worse than those for the SM in both $m_{12}$ and $m_{34}$ distributions.
The total of the $\chi^2/({\rm d.o.f})$ is 73.7/(33), and the probability to be the statistical
fluctuation is $6.1\times 10^{-5}$. The corresponding probability for the SM is 0.19.

From the $m_{12}$ and $m_{34}$ distributions, 
we conclude that the $Z''$ model with $m_{Z''}=60$ GeV is excluded by the ATLAS analysis
because the $Z''$ effects should have been clearly visible in the case.
The case with different value of the coupling $g_{Z''}$ can be easily estimated in the same way.
On the other hand, the $\chi^2$ of the $Z''$ model with $m_{Z''}=80$ GeV is almost the same as 
the one of the SM. The total $\chi^2$ is $\chi^2/({\rm d.o.f})=40.6$, and the corresponding probability is 0.17.  
Thus, the current ATLAS analysis is not sensitive  to the 
$Z''$ model with $m_{Z''}=80$ GeV.
\bigskip

The difference between the SM and $Z''$ model would be more evident
if one looks only at $4\mu$ channel since the $Z''$  only couples to muons, 
although the ATLAS has not provided the separate results.
In Table~\ref{table_m12_4mu}, the numbers of $4\mu$ events 
expected in several $m_{12}$ ranges for the SM and $Z''$ model with 
$m_{Z''}=60$ GeV are listed.
Compared with Table~\ref{table_m12}, $\sigma_{Z'',60}$ in the range $m_{12}=(57,63)$~GeV is much larger.

\begin{table}[h]
\begin{center}
\begin{tabular}{ll||r|rr}
\hline
 $4\mu$ channel  & & $N_{\rm SM}$ & $N_{Z'',60}$ & $\sigma_{Z'',60}$\\
\hline \hline
&$(51, 57)$~GeV & 13.4 & 17.1	& 1.0 \\
$m_{12}$ & $(57,63)$~GeV & 17.4 & 47.3	& 7.2 \\
&$(63,69)$~GeV & 17.5 & 14.1	& -0.8 \\
\hline
\end{tabular}
\end{center}
\caption{Numbers of events in several $m_{12}$ ranges in $4\mu$ channel for 
the SM ($N_{\rm SM}$) and $Z''$ model with $m_{Z''}=60$ GeV ($N_{Z''}$). 
We also show $ \sigma_{Z''}=(N_{Z''}-N_{\rm SM})/\sqrt{N_{\rm SM}}$.}
\label{table_m12_4mu}
\end{table}

For $m_{Z''}=80$ GeV, the $Z''$ model is not constrained by the ATLAS analysis.
In the case of $m_{Z''} \simeq m_{Z}$ or $m_{Z''} > m_{Z}$, 
the off-shell $Z$ boson in the $s$-channel diagram as shown in Figure~\ref{ZpFeyn} 
is dominant in the signal events.
That is the reason why the ATLAS measurement of the $Z$ decays 
to 4 leptons is not sensitive for the heavier $Z''$ boson.

\begin{table}[b]
\begin{center}
\begin{tabular}{l|rr}
\hline
cut & $N_{Z'',60}/N_{\rm SM}$ & $N_{Z'',80}/N_{\rm SM}$ \\
\hline \hline
1. 4 $\mu$ events & 1.21 & 1.02 \\
2-3. $p_{T,\mu}$ and $\Delta R_{\mu,\mu}$ cut & 1.26 & 1.04 \\
4-5. $m_{\mu\mu}$ cuts & 1.28 & 1.05 \\
6. $m_{4\mu}$ in $(80, 100)$~GeV& 1.38 & 0.99 \\
\hline
\end{tabular}
\end{center}
\caption{Ratio of the event numbers in $Z''$ models divided by the SM one
after the successive selection cuts discussed in the text.}
\label{cuts}
\end{table}
We can see this more clearly by the ratios of event numbers 
$N_{Z'',60}/N_{\rm SM}$ and $N_{Z'',80}/N_{\rm SM}$ 
in $4\mu$ channel after the successive selection cuts shown in Table~\ref{cuts}.
The cuts $1-6$ correspond to the ones summarized above. For the cut 1, additionally we require 
$m_{\mu^+\mu^-}>4$ GeV for any combinations of opposite sign di-muons.
The sensitivity to the signal increases as we apply more cuts
in the $Z''$ model with $m_{Z''}=60$ GeV while it decreases 
after the cut 6 in the $Z''$ model with $m_{Z''}=80$~GeV.
It is because the signal events for $m_{Z''}=80$~GeV mostly come from the off-shell region of $Z$ boson.
Consequently, the ATLAS analysis is not directly sensitive to the heavier $Z''$ bosons.

In order to gain sensitivity for the heavier $Z''$ boson, we propose optimized selection cuts:
\begin{itemize}
\item[5'] $m_{4l}>m_Z+ 10$ GeV and reject the Higgs mass region, $|m_{4l} -m_h| > 10$ GeV.
\item[6'] $|m_{34} - m_Z| >5$ GeV.
\end{itemize}
in addition to $p_T$, $\eta$ and $\Delta R$ cuts (cuts $1-4$).
Since the signal events are mainly through $s$-channel off-shell $Z$ boson, 
we reject the contributions through on-shell $Z$ boson as well as 
on-shell Higgs boson by the first criteria (5').
The second criteria (6') is for rejecting $ZZ$ production process,  
which is another SM background, where both $m_{12}$ and $m_{34}$ tend to be close 
to $m_Z$.
On the other hand, in the $Z''$ signal events, $m_{12}$ tends to be $m_{Z''}$, 
but $m_{34}$ does not accumulate on any particular value.
Therefore, it efficiently rejects the $ZZ$ backgrounds while keeping most of the $Z''$ signal.

\begin{figure}[b]
\subfigure[]{\includegraphics*[width=5.334cm]{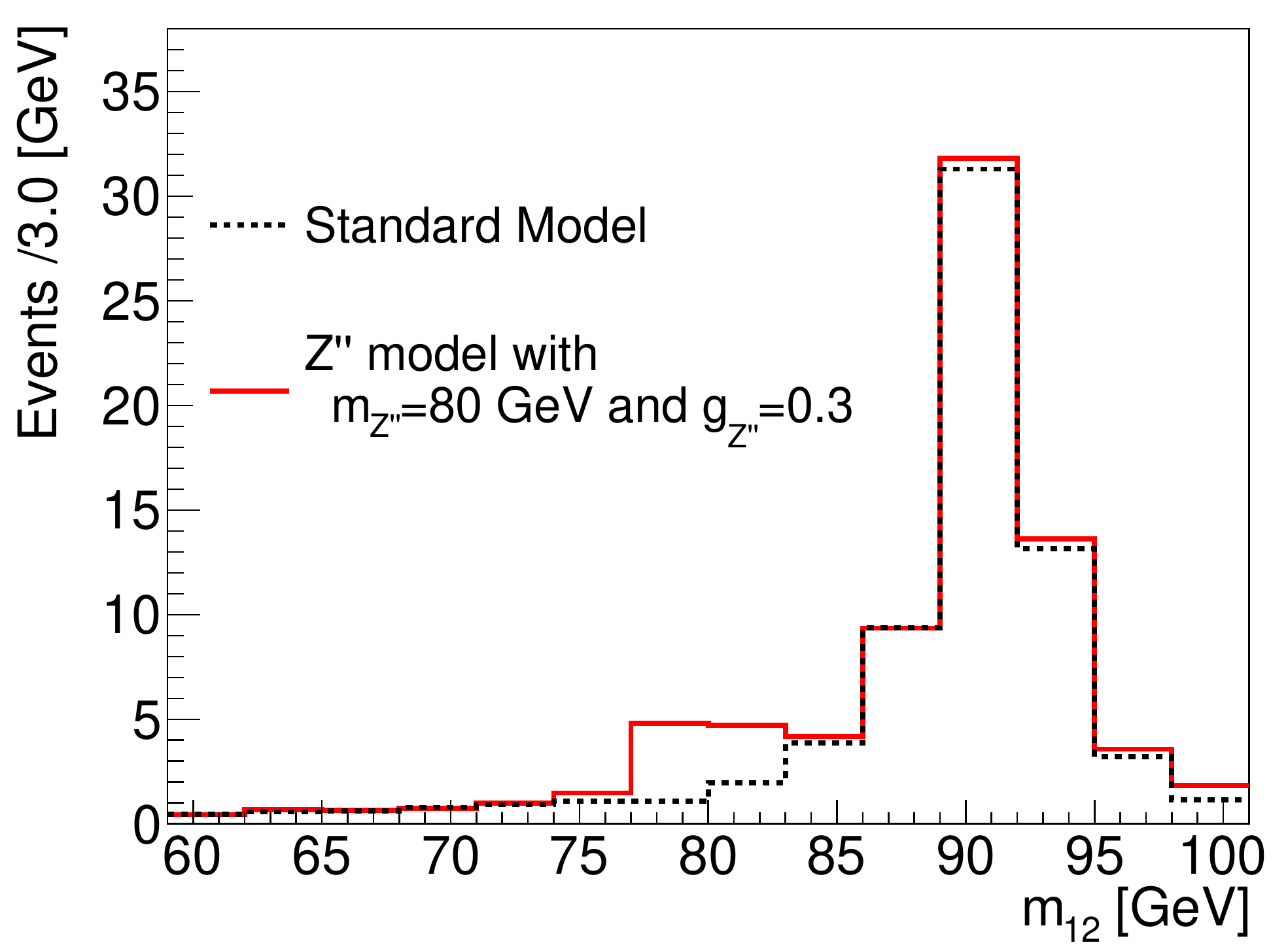}
\label{Zp80_m12_8TeV}}
\subfigure[]{\includegraphics*[width=5.334cm]{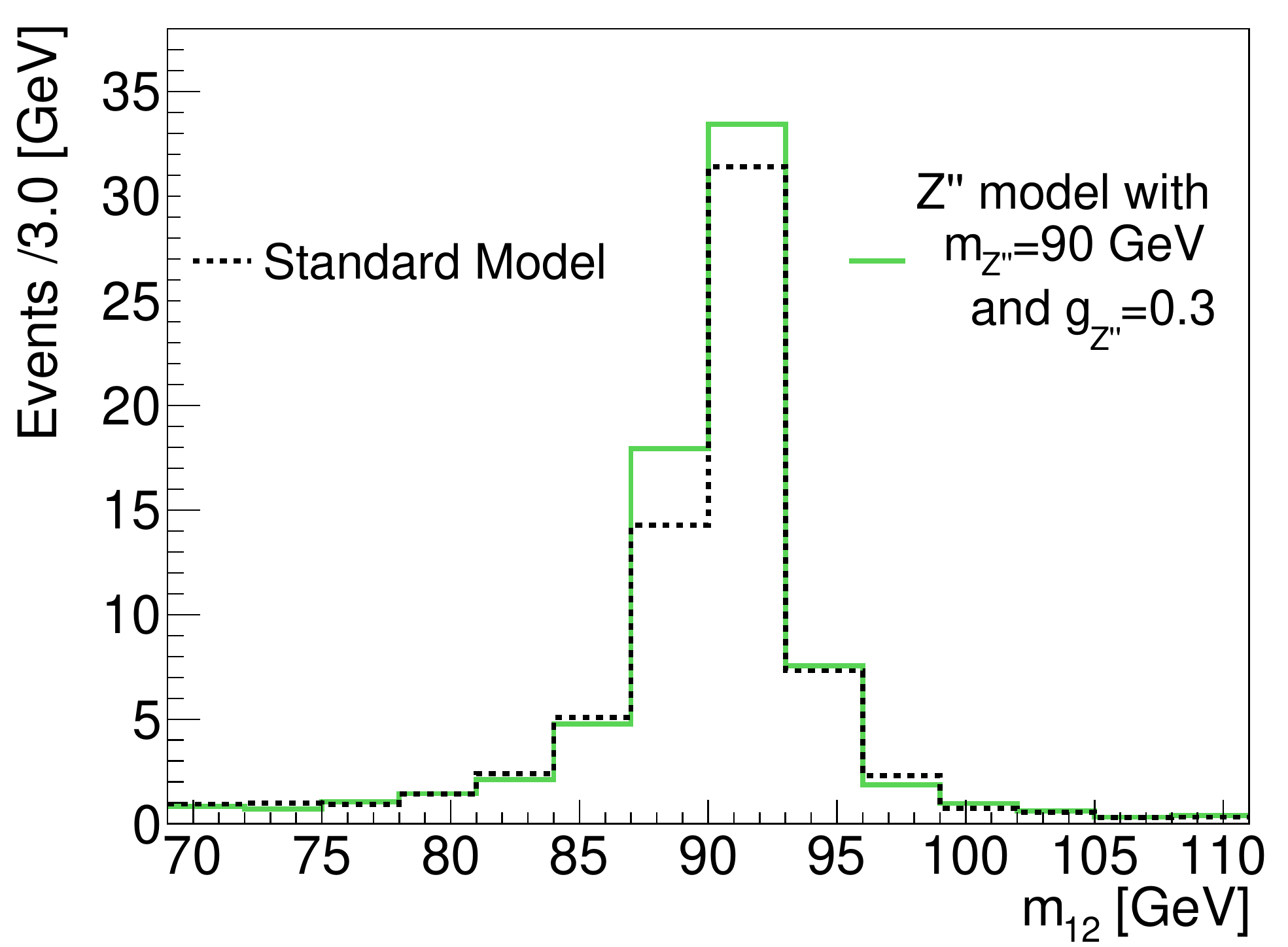}
\label{Zp90_m12_8TeV}}
\subfigure[]{\includegraphics*[width=5.334cm]{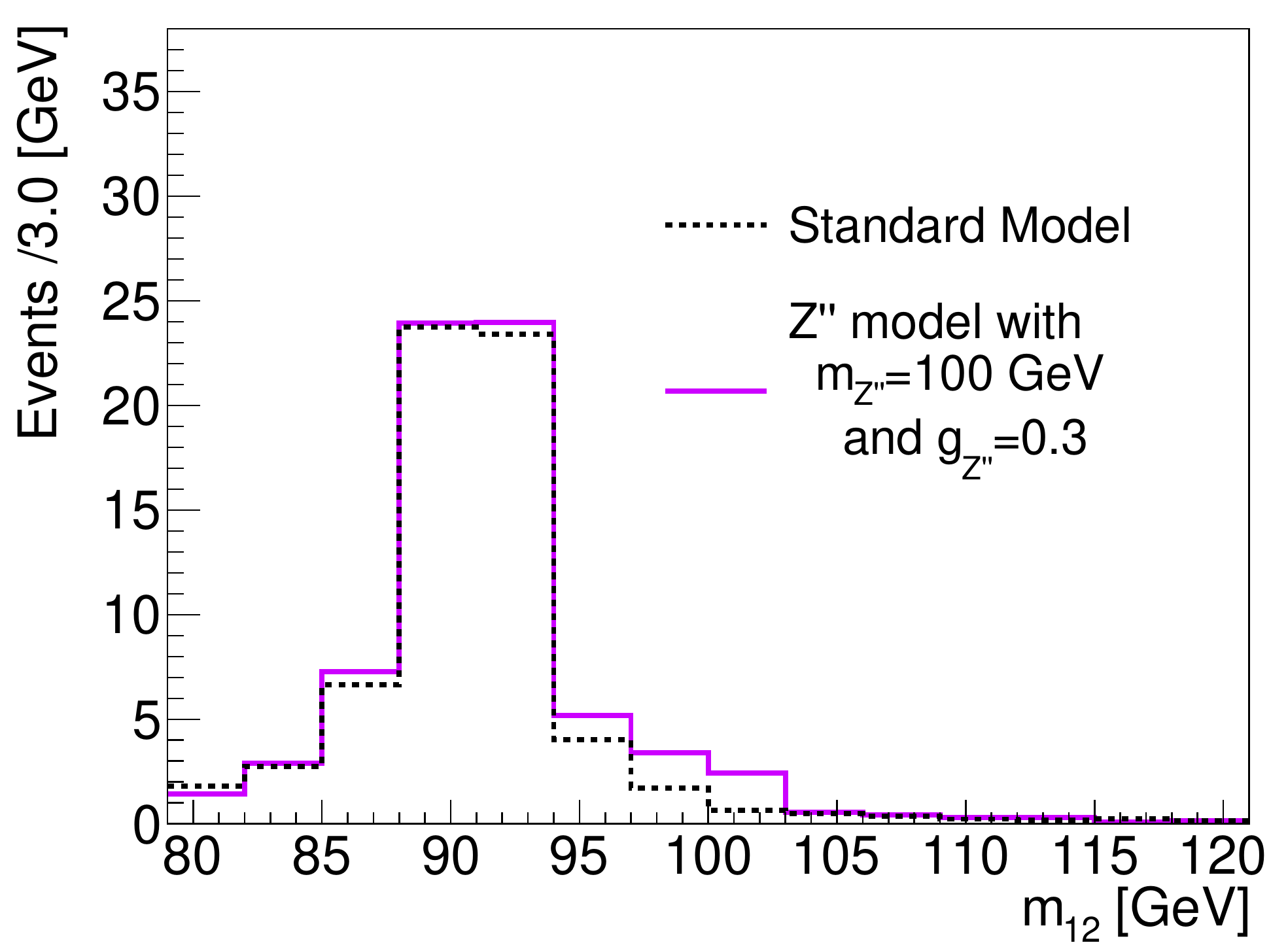}
\label{Zp100_m12_8TeV}}
\caption{The distribution of the di-muon invariant mass $m_{12}$ in $pp\rightarrow 4\mu$
in the SM (dashed line), 
$Z''$ model with $m_{Z''}=80$, 90, and 100~GeV (solid lines, from left to right) after imposing the optimized cuts.
Combined integrated luminosities of 4.6 fb$^{-1}$ at $\sqrt{s}=7$ TeV and 20.7 fb$^{-1}$ at $\sqrt{s}=8$ TeV are assumed.
}
\label{dis_4mu_m12}
\end{figure}

We show the $m_{12}$ distributions in $4\mu$ channel after applying these optimized cuts 
in Figure~\ref{dis_4mu_m12} for  $m_{Z''}=80$, $90$, and $100$ GeV.
For all masses, excesses at $m_{12}=m_{Z''}$ are expected.
Table~\ref{m12_opt} shows the expected numbers in several $m_{12}$ ranges,
the ratio $N_{Z''}/N_{\rm SM}$, and the significance.    

For $m_{Z''}=80$ and $100$~GeV, the $N_{Z''}/N_{\rm SM}$ ratio is very high, 3.1 and 2.5, respectively. The number of event 
in the SM $N_{\rm SM}$  for $m_{12} \sim m_{Z''}$ is very small so that $\sigma$ defined before does not express statistical significance. We estimate the statistical 
significance based on the Poisson distribution with 
the average number of events  $N_{Z''}$  for the bin with $m_{Z''}-3$~GeV $<m_{12}<m_{Z''}+3$~GeV. 
The probability $p$ to have events number in the bin $N \leq N_{\rm SM}^{\rm mode}$ in the 
Poisson distribution with the average of $N_{Z''}$,
where $N^{\rm mode}_{\rm SM}$ is the mode of the Poisson distribution with the average of $N_{\rm SM}$, is 
$1.5\times 10^{-2}$  for $m_{Z''}=80$~GeV 
( 0.07 for $m_{Z''}=100$~GeV). It is not enough to exclude the $m_{Z''}\geq 80$~GeV. 
\footnote{
On the other hand,  $p$ for $N \ge N_{Z''}^{\rm mode}$ in the Poisson distribution with the average of
$N_{\rm SM}$ is $4.7\times 10^{-3}$ for $m_{Z''}=80$~GeV ($0.096$ for $m_{Z''}=100$~GeV).} 
Since $N_{Z''}/N_{\rm SM}$ ratio is very large, the evidence of the $Z''$  should be obtained at 14TeV runs. 
We also checked the possibility to improve those significance 
by using the di-muon invariant mass closest to the hypothetical $m_{Z''}$ value instead of the $m_{12}$.
However, it is not improved since it increases the SM background in the signal region as well.

For $m_{Z''}=90$ GeV, we can see only a small excess over the SM $Z$ boson peak in the $m_{12}$ distribution in Figure~\ref{dis_4mu_m12} (b). The significance is less than 1 
due to the overlapping large SM $Z$ contributions.
\begin{table}[t]
\begin{center}
\begin{tabular}{ll|rr|r|r}
\hline
\multicolumn{2}{l|}{$m_{Z''}=80$~GeV}
& $N_{\rm SM}$ & $N_{Z''}$ & $N_{Z''}/N_{\rm SM}$ & prob. for $N<N_{\rm SM}^{\rm mode}$ in $Z''$ model\\
\hline
&(71,77)~GeV&	2.0		&	2.5 	&	1.2	& -- \\
$m_{12}$ & (77,83)~GeV&	3.1		&	9.5 	&	3.1	& $1.5\times 10^{-2} $ 	\\
&(83,89)~GeV&	13.2	&	13.5 	&	1.0	& -- \\
\hline
\hline
\multicolumn{2}{l|}{$m_{Z''}=90$~GeV}
& $N_{\rm SM}$ & $N_{Z''}$ & $N_{Z''}/N_{\rm SM}$ & $\sigma_{Z''}$\\
\hline
&(81,87)~GeV&	7.5		&	6.9 	&	0.9	& --\\
$m_{12}$ &(87,93)~GeV&	45.7	&	51.4 	&	1.1	&	0.9	\\
&(93,99)~GeV&	9.6		&	9.4 	&	1.0	& -- \\
\hline
\hline
\multicolumn{2}{l|}{$m_{Z''}=100$~GeV}
& $N_{\rm SM}$ & $N_{Z''}$ & $N_{Z''}/N_{\rm SM}$ & prob. for $N<N_{\rm SM}^{\rm mode}$ in $Z''$ model \\
\hline
& (91,97)~GeV&	27.4&	29.2 	&	1.1	& -- \\
$m_{12}$ & (97,103)~GeV&	2.4	&	5.8 	&	2.5	& 0.07 	\\
&(103,109)~GeV&	0.9	&	1.0 	&	1.1	& -- \\
\hline
\end{tabular}
\end{center}
\caption{Numbers in several $m_{12}$ ranges in $4\mu$ channel after applying the optimized cuts
in the SM and in the $Z''$ model for $m_{Z''}=80$, $90$, and $100$ GeV.
We assume combined integrated luminosities of $4.6~{\rm fb}^{-1}$ at $\sqrt{s}=7$ TeV 
and $20.7~{\rm fb}^{-1}$ at $\sqrt{s}=8$ TeV.
}
\label{m12_opt}
\end{table}
%

\subsection{4 lepton channels at $\sqrt{s}=14$ TeV}
 In the previous section, we have shown that the heavy $Z''$ boson ($m_{Z''} >80$~GeV) 
 cannot be excluded by using the $\sqrt{s}=7-8$ TeV run data of the LHC
 due to the limited integrated luminosities. 
In this section, we study the $Z''$ search  at $\sqrt{s}=14$ TeV with the integrated luminosity of 300
fb$^{-1}$ and 3000 fb$^{-1}$ for the same reference points defined in the previous subsection
except the case with $m_{Z''}=60$~GeV, which is already excluded. 
In this section, the leading order results without constant normalization factors are used for the cross sections.
For the  detector simulation, 
we adopt the trigger conditions for the run at $\sqrt{s}=14$ TeV~\cite{LHC14trigger}, as 
shown in Table~\ref{trigger} and implemented in Delphes.
In addition to the $4 \mu$ channel discussed in the previous section, 
we also discuss the channels involving $\tau$-leptons such as $2\mu 2\tau$ and $4\tau$
since $Z''$ also couples to $\tau$-leptons.

\begin{table}[h]
\begin{center}
\begin{tabular}{l|l}
\hline
trigger			&	$p_T$ threshold 			\\
\hline\hline
single muon		&	$p_T^{\mu}>25$ GeV					\\
single tau jet		&	$p_T^{\tau}>150$ GeV				\\
di-muon			&	$p_T^{\mu_1}>13$ GeV,~ $p_T^{\mu_2}>13$ GeV\\
muon-tau jet		&	$p_T^{\mu}>15$ GeV,~ $p_T^{\tau}>40$ GeV	\\
\hline
\end{tabular}
\end{center}
\caption{Trigger conditions relevant for the analysis at $\sqrt{s}=14$ TeV~\cite{LHC14trigger}.
}
\label{trigger}
\end{table}
%

\subsubsection{$pp\rightarrow \mu^+\mu^-\mu^+\mu^-$}
The excess of the $Z''$ signal which lies near $Z$ boson mass in the $m_{12}$ distribution would 
be confirmed at $\sqrt{s}=14$ TeV since the cross section and number of events of 
the $Z''$ model would increase if $Z''$ boson exists for $pp\rightarrow 4\mu$ channels.   
We apply the optimized cuts for the heavier $Z''$ boson 
proposed in the previous section. 

\begin{figure}[h]
\subfigure[]{\includegraphics*[width=5.334cm]{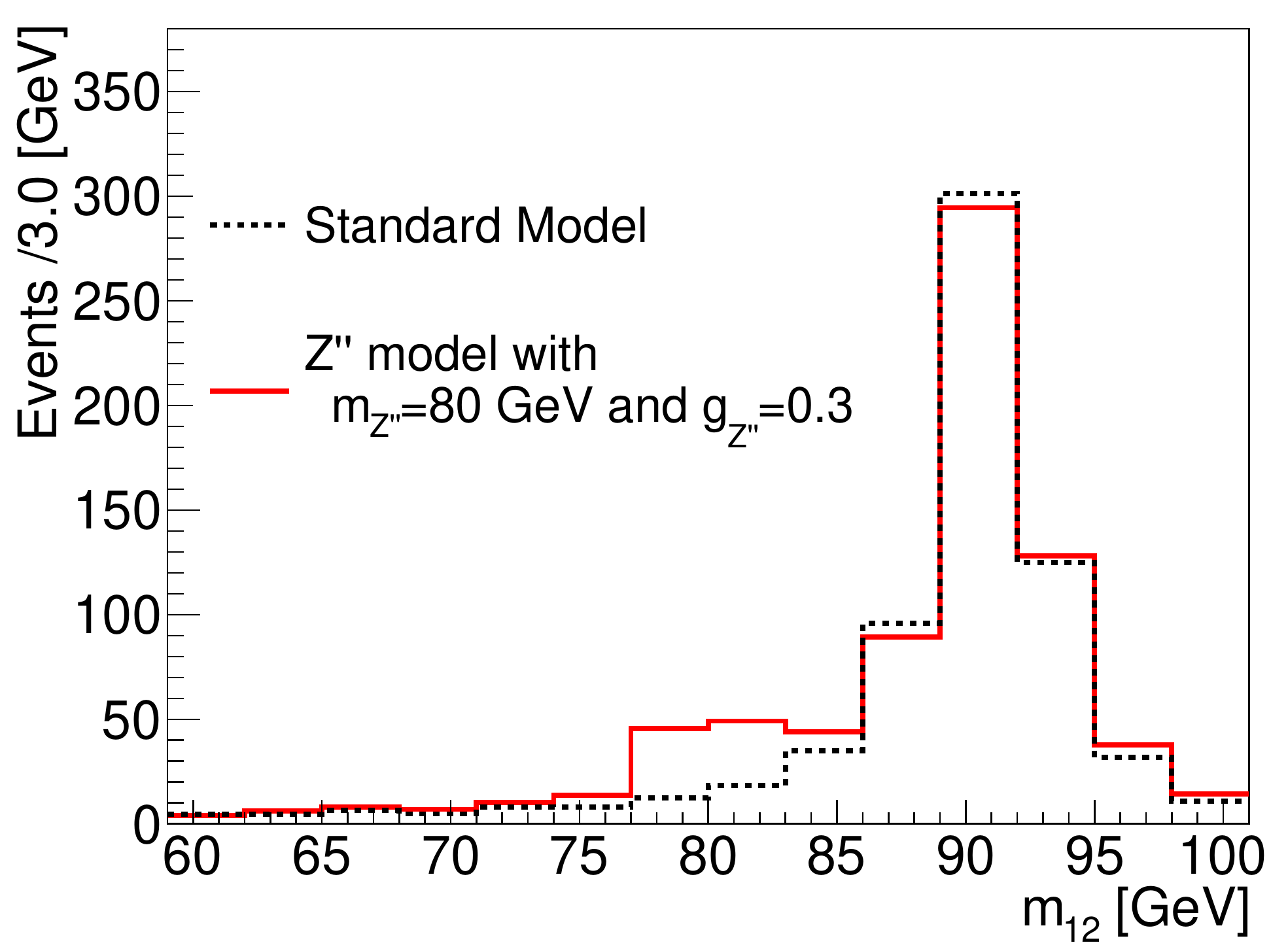}
\label{Zp80_m12_14TeV}}
\subfigure[]{\includegraphics*[width=5.334cm]{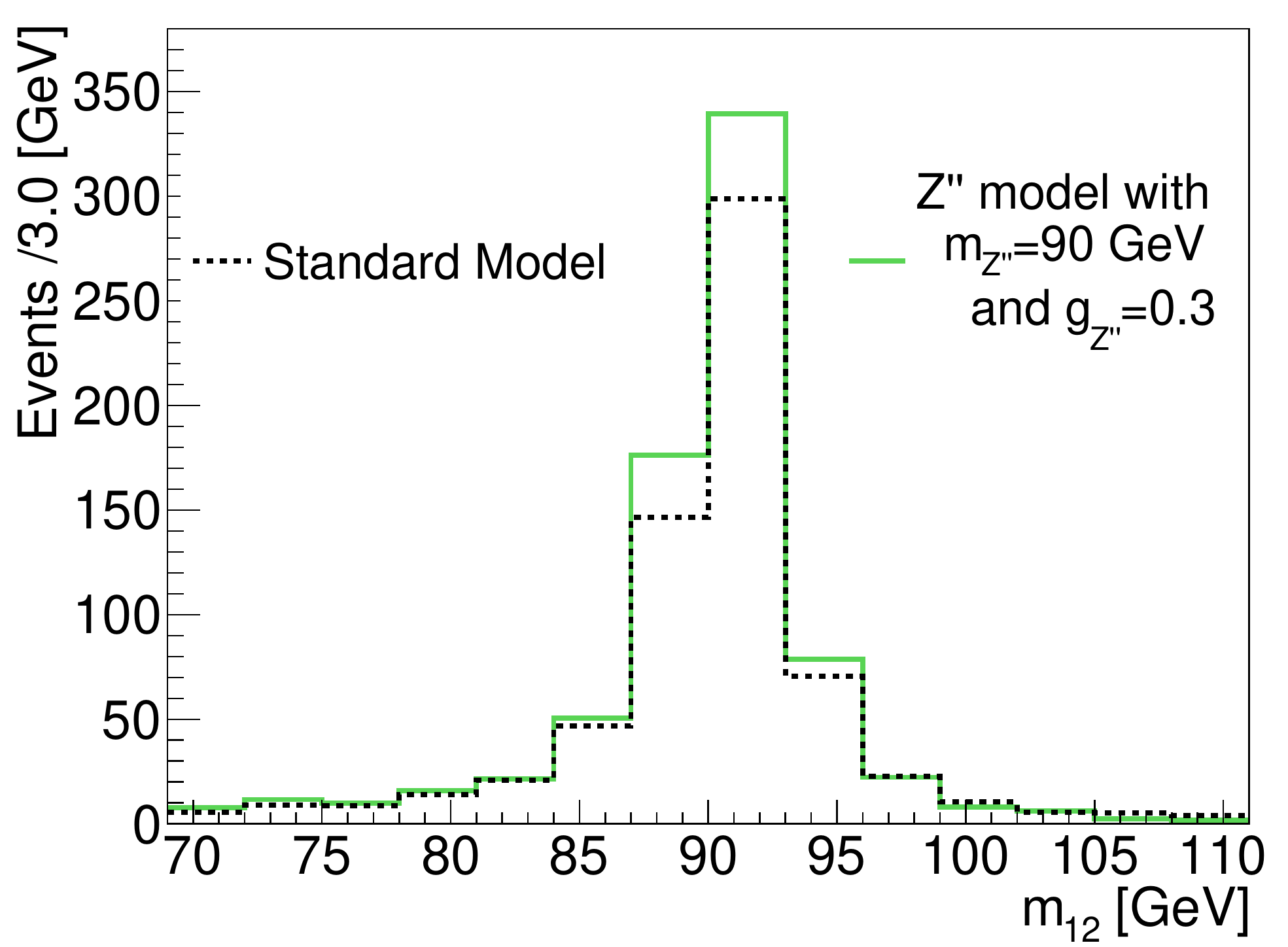}
\label{Zp90_m12_14TeV}}
\subfigure[]{\includegraphics*[width=5.334cm]{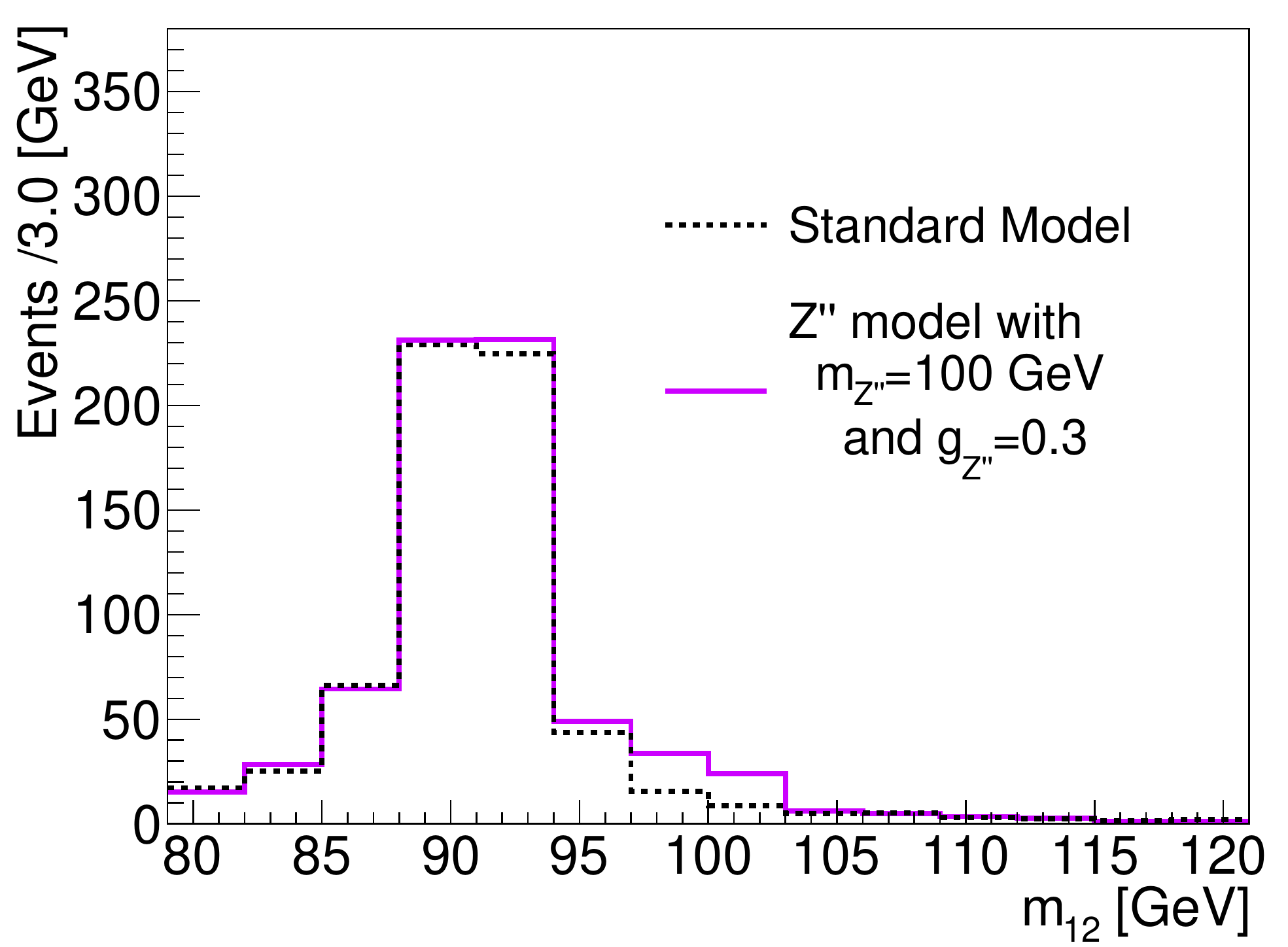}
\label{Zp100_m12_14TeV}}
\caption{The $m_{12}$ distributions in the $4\mu$ channel in the SM (dashed line) 
and $Z''$ model with $m_{Z''}=80$, 90, and $100$ GeV (solid line, from left to right)
after the optimized selection cuts.
The integrated luminosity of 300 fb$^{-1}$ at $\sqrt{s}=14$ TeV is assumed.
}
\label{dis_4mu_m12_14TeV}
\end{figure}

In Figure~\ref{dis_4mu_m12_14TeV}, we show the distributions of the di-muon invariant mass $m_{12}$ for the SM
and the $Z''$ model with $m_{Z''}=80$, 90, and 100~GeV. 
We normalize the distributions for the integrated luminosity of $300~{\rm fb}^{-1}$.
Excesses are more clearly seen in the $Z''$ model with $m_{Z''}=80$ and 100~GeV in the signal region
 $m_{12}\simeq m_{Z''}$, compared with the case at $\sqrt{s}=7-8$ TeV.
Even in the case of $m_{Z''}=90$ GeV, the excess in the region $m_{12}\simeq m_{Z''}$ is statistically significant.
In Table~\ref{m12_opt_14TeV}, the numbers of events around the excesses for the SM $(N_{\rm SM})$, 
for the $Z''$ models ($N_{Z''}$), the ratio $N_{Z''}/N_{\rm SM}$ and $\sigma_{Z''}$ are shown.
The significance $\sigma_{Z'', 80}$ and $\sigma_{Z'', 100}$ exceed 5.
Although we obtain $\sigma_{Z'',90} \sim 3$ for  $m_{Z''}=90$ GeV, 
whole region below $m_{Z''}\leq 100$ GeV will be explored at the high luminosity LHC (HL-LHC), where 
the expected integrated luminosity is around $1000-3000$ fb$^{-1}$, 
provided the cuts and background rate remain the same.

\begin{table}[t]
\begin{center}
\begin{tabular}{ll|rr|r|r}
\hline
\multicolumn{2}{l|}{$m_{Z''}=80$~GeV}
& $N_{\rm SM}$ & $N_{Z''}$ & $N_{Z''}/N_{\rm SM}$ & $\sigma_{Z''}$\\
\hline
&(71, 77) GeV &	16.2	&	23.8	&	1.5	&	1.9		\\
$m_{12}$ & (77, 83) GeV &	30.7	&	94.7	&	3.1	&	11.6	\\
&(83, 89) GeV&	130.8	&	133.3	&	1.0	&	0.2		\\
\hline
\hline
\multicolumn{2}{l|}{$m_{Z''}=90$~GeV}
 & $N_{\rm SM}$ & $N_{Z''}$ & $N_{Z''}/N_{\rm SM}$ & $\sigma_{Z''}$\\
\hline
&(81, 87) GeV&	67.6	&	72.1	&	1.1	&	0.5 \\
$m_{12}$ & (87, 93) GeV&	445.3	&	515.7 	&	1.2	&	3.3	\\
&(93, 99) GeV&	93.3 	&	100.9	&	1.1	&	0.8	\\
\hline
\hline
\multicolumn{2}{l|}{$m_{Z''}=100$~GeV}
& $N_{\rm SM}$ & $N_{Z''}$ & $N_{Z''}/N_{\rm SM}$ & $\sigma_{Z''}$\\
\hline
&(91, 97) GeV&	268.2	&	280.4	&	1.1	&	0.7	\\
$m_{12}$ & (97, 103) GeV&	24.3	&	57.8	&	2.4	&	6.8	\\
&(103, 109) GeV&	10.1	&	11.0	&	1.1	&	0.3\\
\hline
\end{tabular}
\end{center}
\caption{Numbers of events in several $m_{12}$ ranges in $4\mu$-channel 
at $\sqrt{s}=14$ TeV with $\int dt L=300~{\rm fb}^{-1}$ in the SM ($N_{\rm SM}$) and
in the $Z''$ model ($N_{Z''}$) with $m_{Z''}=80$, 90, and 100~GeV 
after applying the optimized cuts.
}
\label{m12_opt_14TeV}
\end{table}
%

\subsubsection{$pp\rightarrow \mu^+\mu^- \tau^+\tau^-$}

In our $Z''$ model, the $Z''$ boson couples to the 2nd and 3rd generation leptons.
In order to test the feature, we need to see the pattern of the couplings of the $Z''$ boson.
One of these interesting processes is $2\mu2\tau$ channel.
To study this channel, we adopt hadronic $\tau$ tagging algorithm of Delphes 
which roughly reproduce   ATLAS and CMS data for $Z\rightarrow \tau^+\tau^-$ channel~\cite{Ovyn:2009tx}. 

For this channel we require the following cuts:

\begin{enumerate}
\item two $\tau$ jets exist satisfying $p_{T,\tau}>20$ GeV and $|\eta_\tau|<2.3$, only hadronically decaying $\tau$'s.
\item two oppositely charged muons exist satisfying $p_{T,\mu}>10$ GeV and $|\eta_\mu|<2.7$, the two muons are well separated as $\Delta R >0.1$.
\item requiring the invariant mass cut for the two $\tau$'s, $m_{\tau\tau}>120$ GeV, where 
we adopt the collinear approximation for the $\tau$ momentum reconstruction, 
that is, the neutrino momentum from $\tau$ decay is assumed to
be parallel to the $\tau$ jet direction.
\end{enumerate}

The 1st and 2nd requirements select events which have $2\mu$ and $2\tau$. 
The 3rd cut effectively rejects the SM $ZZ$ backgrounds. 
It is because the signal matrix element is not enhanced at
$m_{\tau\tau} \sim m_{Z}$ nor $m_{Z''}$ once we require $m_{\mu\mu} \sim m_{Z''}$.
On the other hand, in the SM $ZZ$ background
both $m_{\mu\mu}$  and $m_{\tau\tau}$ are enhanced at  $m_{Z}$.
We found that the collinear approximation for the $\tau$ reconstruction is not good enough to
reproduce the $Z''$ mass from the di-tau invariant mass. 
Nevertheless, we found it useful to reject the SM background.

\begin{figure}[h]
\subfigure[]{\includegraphics*[width=5.334cm]{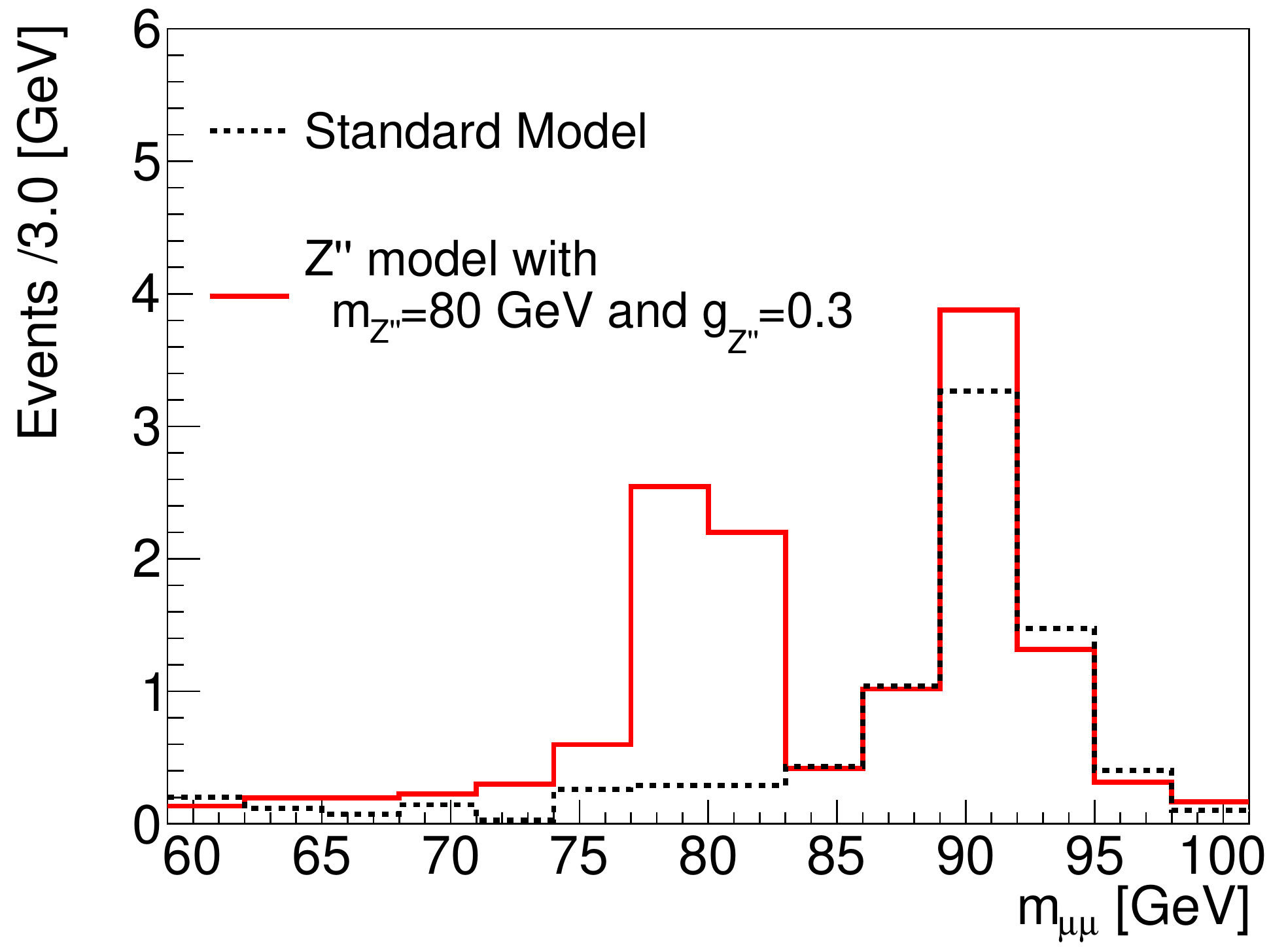}
\label{2m2t_80}}
\subfigure[]{\includegraphics*[width=5.334cm]{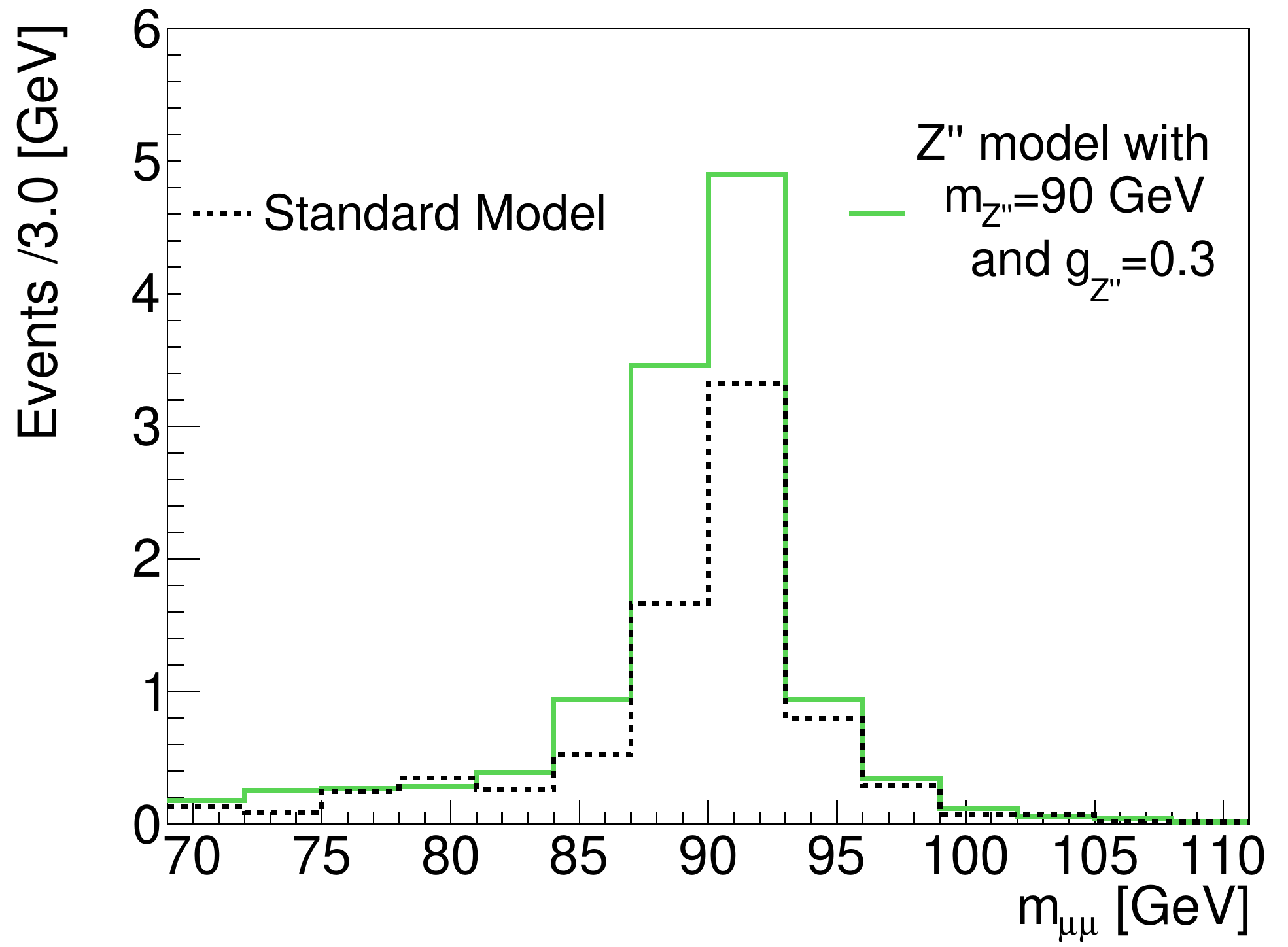}
\label{2m2t_90}}
\subfigure[]{\includegraphics*[width=5.334cm]{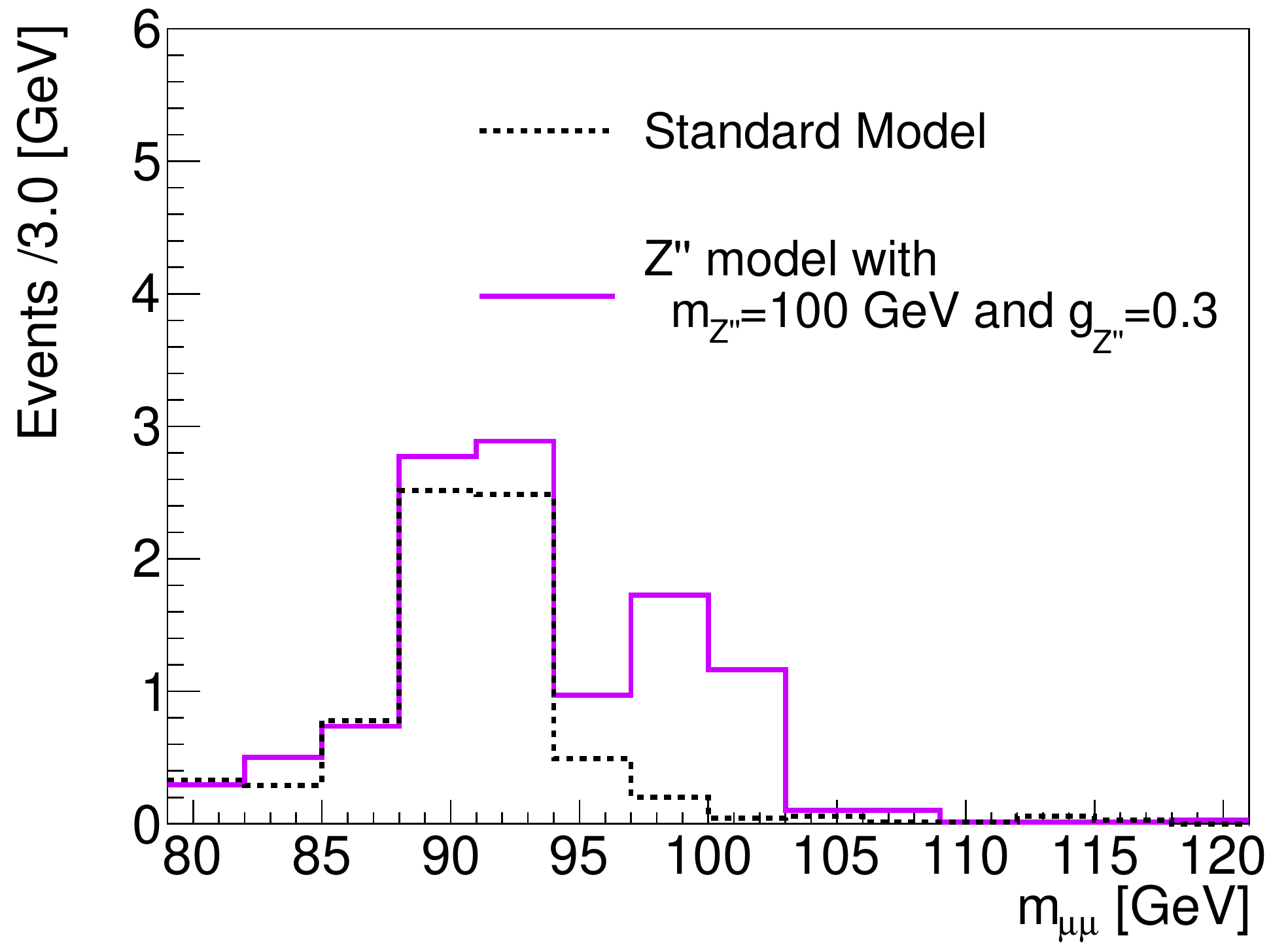}
\label{2m2t_100}}
\caption{
The  $(m_{\mu\mu})$ distributions in the $2\mu 2\tau$ channel at $\sqrt{s}=14$ TeV 
for the SM (dashed line) and for the $Z''$ model with $m_{Z''}=80,~90$,  and 100~GeV (solid lines, from left to right).
The integrated luminosity of 300 fb$^{-1}$ is assumed. 
}
\label{2m2t_14TeV}
\end{figure}

In $4\mu$ channel there are two possible combinations to pair the muons.
We have primarily used the $m_{12}$, which is the lepton pair closer to $m_Z$, for the $Z''$ boson search.
By contrast, $2\mu2\tau$ channel has no such combinatorial problem.
In Figure~\ref{2m2t_14TeV}, we show the di-muon invariant mass ($m_{\mu\mu}$) distributions 
for the SM (dashed line) and $Z''$ model with $m_{Z''}=80$,~90, and 100~GeV (solid lines), from left to right panels, respectively. The normalizations are for the integrated luminosity of $300~{\rm fb}^{-1}$. 

\begin{table}[t]
\begin{center}
\begin{tabular}{ll|rr|r|r}
\hline
\multicolumn{2}{l|}{$m_{Z''}=80$~GeV}
& $N_{\rm SM}$ & $N_{Z''}$ & $N_{Z''}/N_{\rm SM}$ & $\int dt L$ for discovery (fb$^{-1}$) \\
\hline
&(71, 77) GeV &	0.3	&	0.9	&	3.1	&		\\
$m_{\mu\mu}$ & (77, 83) GeV &	0.6	&	4.8	&	8.2	& $>500$\\
&(83, 89) GeV&	1.5	&	1.4	&	1.0	&		\\
\hline
\hline
\multicolumn{2}{l|}{$m_{Z''}=90$~GeV}
& $N_{\rm SM}$ & $N_{Z''}$ & $N_{Z''}/N_{\rm SM}$ & $\int dt L$ for discovery(fb$^{-1}$)\\
\hline
&(81, 87) GeV&	0.8	&	1.3	&	1.7	&		\\
$m_{\mu\mu}$& (87, 93) GeV&	5.0	&	8.4	&	1.7	&	$>290$0	\\
&(93, 99) GeV&	1.1	&	1.3	&	1.2	&		\\
\hline
\hline
\multicolumn{2}{l|}{$m_{Z''}=100$~GeV}
& $N_{\rm SM}$ & $N_{Z''}$ & $N_{Z''}/N_{\rm SM}$ & $\int dt L$ for discovery(fb$^{-1}$)\\
\hline
&(91, 97) GeV&	3.0		&	3.7	&	1.3		&	\\
$m_{\mu\mu}$ & (97, 103) GeV&	0.3		&	2.9	&	11.8	& $>730$	\\
&(103, 109) GeV&	0.07	&	0.2	&	2.9		&		\\
\hline
\end{tabular}
\end{center}
\caption{Number of events in several $m_{\mu\mu}$ ranges in $2\mu 2\tau$ channel
at $\sqrt{s}=14$ TeV with $\int dt L=300~{\rm fb}^{-1}$ in the SM and the 
$Z''$ model with $m_{Z''}=80$,~90, and ~100 GeV.
}
\label{table_2m2t_14TeV}
\end{table}

Table~\ref{table_2m2t_14TeV} shows event numbers in several $m_{\mu\mu}$ ranges 
around the excess for the $Z''$ models ($N_{Z''}$) together with those for the SM $(N_{\rm SM})$, 
the ratio $N_{Z''}/N_{\rm SM}$, and significance expressed by the required integrated luminosity for 
the discovery, which is defined as the integrated luminosity where the probability to have number of events 
in the signal bin $N >N_{Z''} $ is less than $10^{-5}$  for the Poisson distribution with the average of $N_{\rm SM}$. 
The $N_{Z''}/N_{\rm SM}$ is large enough at $m_{\mu\mu}\sim m_{Z''}$  for both cases of $m_{Z''}=80$ and 100 GeV, 
 while it is only around 1.7 for  $m_{Z''}=90$ GeV.
The $Z''$ effects will be observed  for $m_{Z''}=80$~GeV with the luminosity less than 300 fb$^{-1}$.
In the case of $m_{Z''}=100$~GeV the number of signal events is small, but 
more data at the HL-LHC would strengthen the signal observation. 
In the case of $m_{Z''}=90$ GeV, the significance is smaller, 
however, it would be possible to observe the definite signal 
once we collect more data with the integrated luminosity of $3000~{\rm fb}^{-1}$ at the HL-LHC 
assuming the acceptance of leptons and $\tau$ jets unchanged. 

\section{Conclusion}
New particles with the mass of the order of the EW scale with a significant coupling 
to the muon sector can accommodate the muon g-2 
anomaly. The LHC would be an important experiment to search such particles directly 
because of the high luminosity and cleanness of the muon signature.

In this paper, we have considered the $L_\mu-L_\tau$ gauge symmetry 
as one of the solutions to explain the anomaly of muon g-2.
We have explicitly shown that the $Z''$ gauge boson of the EW scale mass 
explains the anomaly of muon g-2. We have also identified the parameter space 
where the $Z''$ model is consistent with the EW precision measurements.
We have considered the LHC phenomenology for several reference model points in the preferred parameter space. 
 
The $Z''$ model contribution to $Z\rightarrow 4\mu$ is large for the relatively light $Z''$ boson
since a $Z$ boson can decay into the $Z''$ boson.
Therefore, we have closely checked the measurement of $Z$ boson decay to 
4 leptons ($4e,~4\mu,~2e 2\mu$) at the ATLAS experiment.
We conclude that the ATLAS result has already excluded the $Z''$ model 
with $m_{Z''}=60$ GeV for $g_{Z''}=0.3$.
The ATLAS analysis is not sensitive yet  to  $Z''$ bosons with mass above  $\sim$ 80GeV. 
We have proposed an analysis  in $pp\rightarrow 4\mu$ channel 
sensitive to the heavier $Z''$ bosons, 
and have shown that the data at $\sqrt{s}=7-8$ TeV should have some sensitivity to the $Z''$ boson 
with $m_{Z''}=80$ and 100~GeV for $g_{Z''}=0.3$.

Moreover, we have shown that LHC data  at the 14 TeV LHC with 300 fb$^{-1}$ 
would be enough to observe the clear $Z''$ boson signal in $4\mu$ channel 
with $m_{Z''}=80$ and 100~GeV for $g_{Z''}=0.3$.
Even in the case of $m_{Z''}=90$ GeV, the integrated luminosity of 3000 fb$^{-1}$ 
would reveal the $Z''$ model.
Therefore, the current and future LHC data in the $4\mu$ final state 
will provide the opportunity to explore the whole region of the 
$Z''$ model parameter space relevant to the muon g-2 anomaly.

In order to probe the $Z''$ model, we should observe the $Z''$ effects 
not only in the $4\mu$ final state but also in the channels involving $\tau$ leptons such as the $2\mu 2\tau$ state 
since it is the important feature that the $Z''$ boson 
only couples to the 2nd and 3rd generation leptons.
We have shown that the $Z''\tau^+\tau^-$ interaction
would be probed in the $2\mu 2\tau$ final state
with the LHC data of the integrated luminosity $3000$ fb$^{-1}$ at $\sqrt{s}=14$~TeV 
for the preferable parameter region of the $Z''$ model. 
Future LHC data  are crucial to test the new physics models 
responsible for the muon g-2 anomaly.

\section*{Acknowledgments}
The authors acknowledge the Yukawa Institute for Theoretical Physics at Kyoto University, 
where this work was initiated during the YITP workshop 
``LHC vs Beyond the Standard Model --Frontier of particle physics (YITP-W-12-21)'' held in March 19th-25th 2013 for
hospitality, and they also thank participants of the workshop for the active discussions.
The work is supported in part by a JSPS Research Fellowship for Young Scientists (K.H.),
World Premier International Research Center Initiative (WPI Initiative), MEXT, Japan
(K.H. and M.M.N.), and Grants-in-Aid for Scientific Research from the Ministry of
Education, Science, Sports, and Culture (MEXT), Japan (No. 23104006 for M.M.N. and
No. 22224003 for K.T.). MT wishes to thank the STFC for support from grant number ST/J002798/1
and Feng Luo for helpful discussions.

\setcounter{equation}{0}
\setcounter{footnote}{1}

\addcontentsline{toc}{section}{Appendix: Passarino-Veltman functions}

\newpage
\appendix

\section{Neutrino mixing and constraint from the washout of the baryon asymmetry}
In this Appendix, we discuss the neutrino mixing in the $L_{\mu}-L_{\tau}$
gauged theory.
We show that the observed neutrino mixing can be explained with an aid
of three right-handed neutrinos. We also discuss the constraint from the
washout of the baryon asymmetry in the early universe, and find that the masses 
of the right-handed neutrinos are bounded from above. 

\subsection{Neutrino mixing}
In addition to the SM fields, we introduce three
right-handed neutrinos $\nu_{iR}(i=e,~\mu,~\tau)$, in order to explain the observed neutrino
mixing.
We assume that they have $L_{\mu}-L_{\tau}$ charges of $0,+1,-1$,
respectively.
We also assume that the $L_{\mu}-L_{\tau}$ gauge symmetry is broken by
a condensation of a scalar field $\sigma$ with a unit $L_{\mu}-L_{\tau}$
charge.
The charge assignments of various fields are summarized in Table~\ref{charge-all}.

\begin{table}[h]
\begin{center}
\begin{tabular}{|c|c|c|c|c|c|c|c|c|c|c|c|}
\hline
particle& $L_1$ & $L_2$ & $L_3$ & $(e_R)^c$ &
$(\mu_R)^c$ & $(\tau_R)^c$ & $(\nu_{eR})^c$ &$(\nu_{\mu R})^c$&$(\nu_{\tau R})^c$&$\sigma$&others\\
\hline \hline
charge &0& +1 & -1 &0&-1 &+1 &0&-1&+1&+1&0\\
\hline
\end{tabular}
\end{center}
\caption{Charges assignments under the $L_\mu-L_\tau$ gauge
 symmetry. All fermion fields are written
in left-handed basis.}
\label{charge-all}
\end{table}

From the charge assignments, renormalizable terms in a Lagrangian which
contribute to the lepton masses are given by
\begin{eqnarray}
\label{eq:mass terms}
 {\cal L} &= &
H^c\left(
y_e L_1 \left(e_R\right)^c +
y_\mu L_2 \left(\mu_R\right)^c+
y_\tau L_3 \left(\tau_R\right)^c
\right)
+H\left(
\lambda_1 L_1 \left(\nu_{eR}\right)^c +
\lambda_2 L_2 \left(\nu_{\mu R}\right)^c +
\lambda_3 L_3 \left(\nu_{\tau R}\right)^c
\right)\nonumber\\
&&
+M_{ee} \nu_{eR} \nu_{eR}+ M_{\mu\tau} \nu_{\mu R} \nu_{\tau R} + \lambda'_{e\mu} \sigma^\dag \nu_{eR} \nu_{\mu R} 
+ \lambda'_{e\tau}
\sigma \nu_{eR} \nu_{\tau R} + {\rm h.c.}.
\end{eqnarray}
Here, $y_e,~y_\mu,~y_\tau$ are the Yukawa couplings of the charged leptons
and not related to the neutrino mass.
The neutrino mass is determined from Yukawa couplings $\lambda_i~(i =
1,2,3)$, Majorana masses $M_{ee}$ and $M_{\mu\tau}$, and Yukawa couplings
$\lambda'_{e\mu}$ and $\lambda'_{e\tau}$.

Note that the mass terms between the left and right-handed neutrinos are
diagonal. Therefore, the neutrino mixing is obtained by mixing among
the right-handed neutrinos. 
If the Majorana masses $M_{ee},~M_{\mu\tau},~\lambda'_{e\mu} \vev{\sigma}$ and
$\lambda'_{e\tau} \vev{\sigma}$ are of the same order, the seesaw
mechanism~\cite{seesaw} provides the observed order one neutrino mixing.
From the seesaw formula, a relation between the parameters is given by
\begin{eqnarray}
\label{eq:seesaw}
 (\Delta m^2 )^{1/2} \sim \frac{\lambda^2 v^2}{M} \sim 10^{-12} - 10^{-11}
 ~{\rm GeV},
\end{eqnarray}
where $v \simeq 174$ GeV is the vacuum expectation value of the SM 
Higgs, and $\Delta m^2$ is the difference between the mass squared
of the left-handed neutrinos.
$\lambda$ and $M$ denote $\lambda_i~(i = 1,2,3)$ and
$M_{ee},~M_{\mu\tau},~\lambda'_{e\mu} \vev{\sigma}$ and $\lambda'_{e\tau} \vev{\sigma}$ collectively.

\subsection{Washout of the baryon asymmetry}
Interactions given by Eq.~(\ref{eq:mass terms}) break the lepton
symmetry.
On the other hand, $B+L$ symmetry is broken by the anomaly against the $SU(2)$ gauge
interaction, whose effect is
efficient at the early universe by the sphaleron process in the finite temperature~\cite{Kuzmin:1985mm}.
Therefore, the baryon asymmetry is washed out if both effects are
important simultaneously.
Let us calculate a condition such that the washout does not occur.

First of all, the sphaleron process is efficient only at the temperature
above the EW scale. Therefore, if the baryon asymmetry is generated
below the EW scale, the washout does not occur.
In the following, we assume that the baryon asymmetry is produced above
the EW scale and calculate the constraint on the parameters in
Eq.~(\ref{eq:mass terms}).

Let us consider two possibilities in which the washout does not occur.
\begin{enumerate}
 \item $\lambda_i$ is small,
 \item $M_{ee}$ and $M_{\mu\tau}$ are small.
\end{enumerate}
If any of the two conditions are satisfied, the lepton number is
effectively conserved.
Therefore, one should adopt the weakest condition among them.
Let us discuss the two cases in detail.

\subsubsection*{$\lambda_i$ is small}
\begin{figure}[h]
\centerline{
\includegraphics[width=7cm]{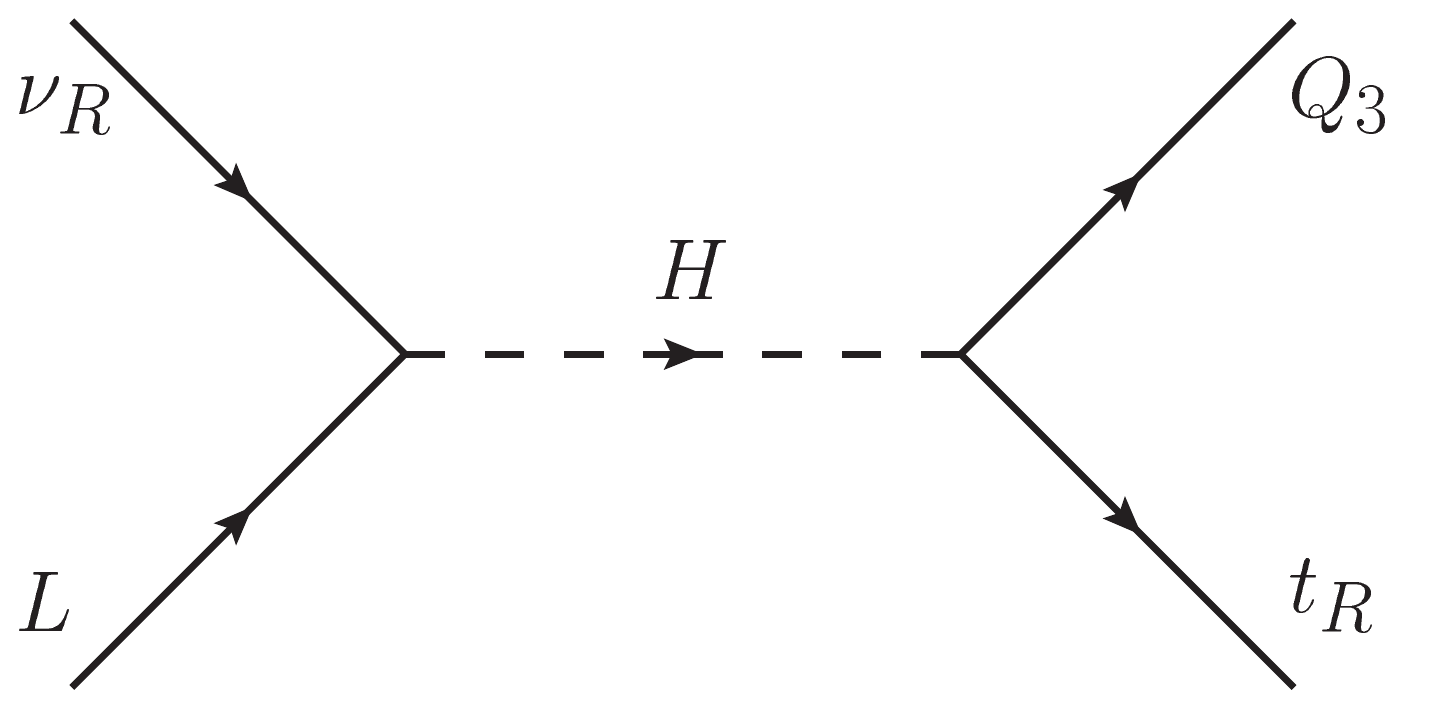}
}
\caption{Feynman diagram for the lepton number violating interaction by $\lambda$.}
\label{fig:lambda}
\end{figure}

In the limit $\lambda_i=0$, the lepton symmetry is restored for each flavors. Therefore,
if the interaction by $\lambda_i$ is inefficient, the washout of the
baryon asymmetry does not occur.
The most efficient interaction is shown in Figure~\ref{fig:lambda} and 
its rate is given by
\begin{eqnarray}
 \vev{\sigma n v} \simeq \frac{\lambda_i^2 y_t^2}{8\pi} T,
\end{eqnarray}
where $\sigma$, $n$, $v$, $y_t$ are the cross section of the process, the
number density of related particles, the velocity of related
particles, and the Yukawa coupling of the top quark, respectively. $\vev{\cdots}$ denotes the thermal average.
 By requiring that the rate
is smaller than the Hubble scale for $T\gsim10^2$ GeV, we obtain the bound
\begin{eqnarray}
\label{eq:constraint-l}
 \lambda_i \lsim 10^{-7}.
\end{eqnarray}

\subsubsection*{$M_{ee}$ and $M_{\mu\tau}$ are small}
\begin{figure}[h]
\centerline{
\includegraphics[width=7cm]{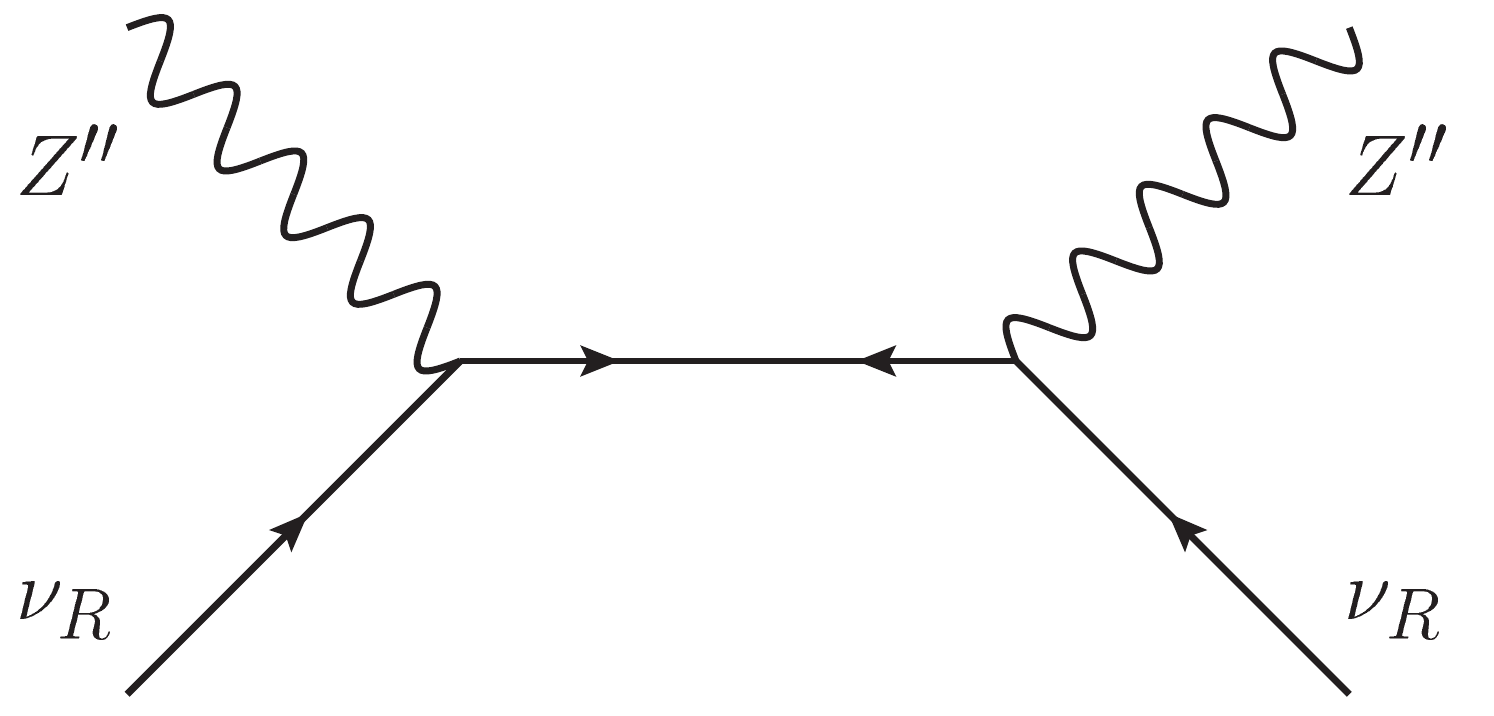}
}
\caption{Feynman diagram for the lepton number violating interaction.}
\label{fig:M}
\end{figure}

If both Majorana masses vanish, $L_e- L_{\mu}-L_\tau $ symmetry is restored.
The most efficient interaction which induces the symmetry violation by
the Majorana masses is shown in Figure~\ref{fig:M}. Its rate is given by
\begin{eqnarray}
 \vev{\sigma n v} \sim \frac{g'^4}{8\pi} M.
\end{eqnarray}
This rate is smaller than the Hubble scale for $T\gsim 10^2$ GeV if 
\begin{eqnarray}
\label{eq:constraint-M}
 M \lsim 10^{-11}~{\rm GeV}~ (\frac{g'}{0.3})^{-4}.
\end{eqnarray}

From the relation (\ref{eq:seesaw}), one can see that the condition
(\ref{eq:constraint-M}) is severer than the condition
(\ref{eq:constraint-l}).
Therefore, it is enough to satisfy the condition (\ref{eq:constraint-l})
in order for the washout not to occur.
%
With the relation (\ref{eq:seesaw}), the condition is
interpreted as
\begin{eqnarray}
 \lambda_i &\lsim& 10^{-7}\nonumber\\
 M_{ee},~M_{\mu\tau}  &\lsim& 10^1~{\rm GeV}\nonumber\\
 \lambda_{e\mu}',~\lambda_{e\tau}' &\lsim& 10^{-1} \frac{\vev{\sigma}}{100~{\rm GeV}}
\end{eqnarray}

\begin{figure}[h]
\centerline{
\includegraphics[width=7cm]{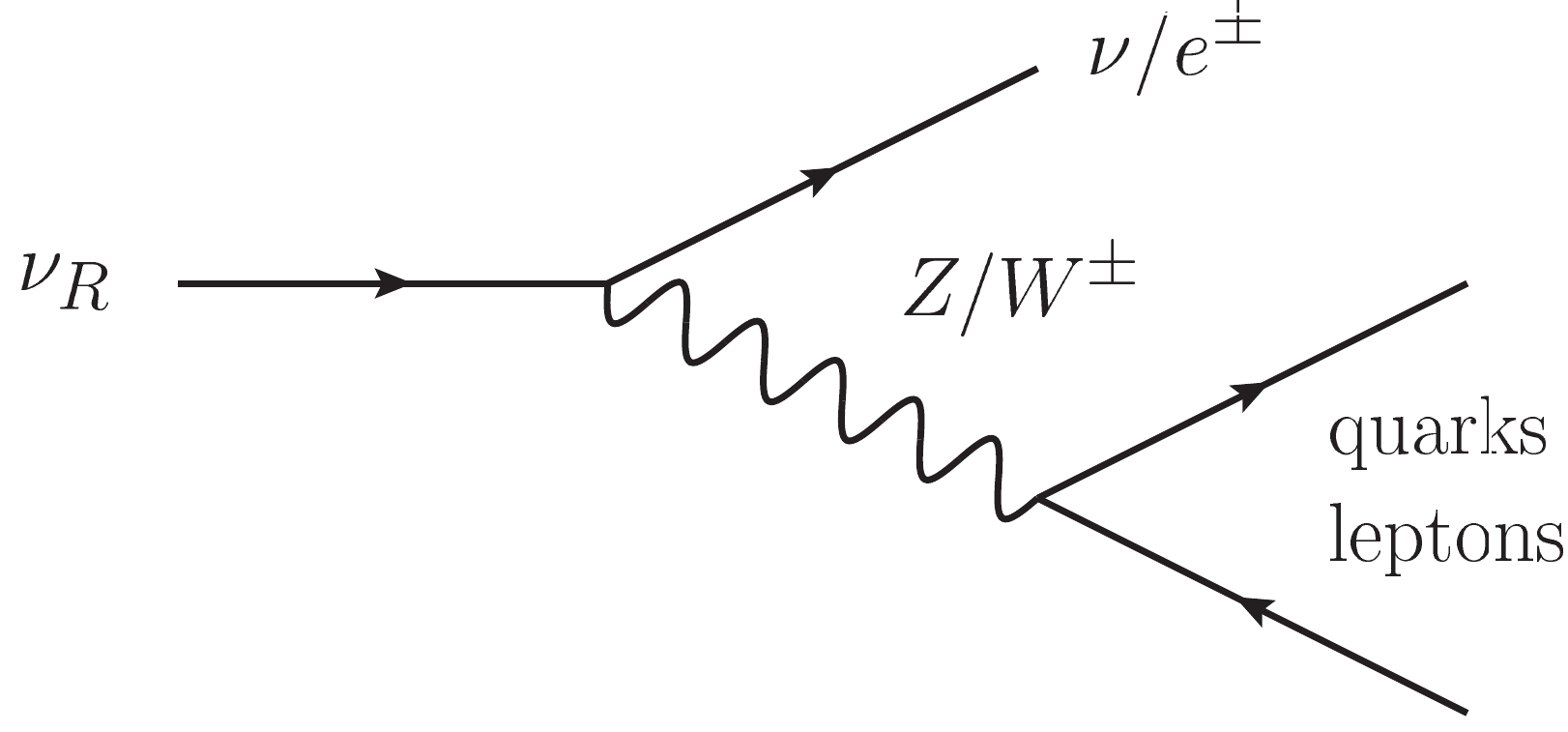}
}
\caption{Feynman diagram for the decay of the right-handed neutrinos.
}
\label{fig:decay of N}
\end{figure}

Since the right-handed neutrinos are light and weakly coupled, it is
necessary to consider whether they are long-lived. If they are
long-lived, they might over-close the universe, or destroy the success of
the big-bang nucleosynthesis (BBN).
The most important decay channel is given by the diagram shown in
Figure.~\ref{fig:decay of N}. Here, we have assumed that $\sigma$ is heavier
than the right-handed neutrinos and hence the decay mode $N\rightarrow \sigma \nu$ is closed.
The decay rate is given by
\begin{eqnarray}
 \Gamma \simeq \frac{M}{128\pi^3} \frac{M^2}{v^2}  \lambda^2.
\end{eqnarray}
The decay of the right-handed neutrinos is efficient around the temperature
\begin{eqnarray}
 T\sim 0.1~{\rm GeV}~\frac{\lambda}{10^{-7}}(\frac{M}{10~{\rm GeV}})^{3/2}.
\end{eqnarray}
Therefore, the right-handed neutrinos decay before the BBN begins and
does not affect it.

\section{Passarino-Veltman functions}
Passarino-Veltman functions~\cite{'tHooft:1978xw} are defined by
\begin{eqnarray}
A(A)&=&
16\pi^2\mu^{2\epsilon}\int
\frac{d^n k}{i(2\pi)^n}\frac{1}{k^2-m_A^2+i\epsilon},
\\
B_0(A,B;p) &=&
16\pi^2\mu^{2\epsilon}\int
\frac{d^nk}{i(2\pi)^n}\frac{1}{\left[k^2-m_A^2+i\epsilon\right]
\left[(k+p)^2-m_B^2+i \epsilon\right]},\nonumber
\\
p^\mu B_1(A,B;p) &=& 16\pi^2 \mu^{2\epsilon}
\int \frac{d^n k}{i(2\pi)^n}\frac{k^\mu}
{\left[k^2-m_A^2+i\epsilon\right]\left[(k+p)^2-m_B^2+i\epsilon\right]},
\nonumber
\\
p^\mu p^\nu B_{21}(A,B;p)&+&g^{\mu \nu}B_{22}(A,B;p)
\nonumber \\
&=& 16\pi^2 \mu^{2\epsilon} \int
\frac{d^n k}{i(2\pi)^n} \frac{k^\mu k^\nu}
{\left[k^2-m_A^2+i\epsilon\right]\left[(k+p)^2-m_B^2+i\epsilon\right]},
\end{eqnarray}
\begin{eqnarray}
&& C_0(A,B,C;p_1,p_2)\nonumber \\
&=&
16\pi^2 \mu^{2\epsilon}
\int \frac{d^n k}{i(2\pi)^n}\frac{1}
{[k^2-m_A^2+i\epsilon][(k+p_1)^2-m_B^2+i\epsilon][(k+p_1+p_2)^2-m_C^2+i\epsilon]},
\nonumber 
\\
&&\left(
p_1^\mu C_{11}+p_2^\mu C_{12}
\right)(A,B,C;p_1,p_2) \nonumber \\
&=& 16\pi^2 \mu^{2\epsilon}
\int \frac{d^n k}{i(2\pi)^n}\frac{k^\mu}
{[k^2-m_A^2+i\epsilon][(k+p_1)^2-m_B^2+i\epsilon][(k+p_1+p_2)^2-m_C^2+i\epsilon]},
\nonumber 
\\
&&\left\{(p_1^\mu p_1^\nu C_{21}
+p_2^\mu p_2^\nu C_{22}+(p_1^\mu p_2^\nu+p_1^\nu p_2^\mu)C_{23} +g^{\mu \nu}C_{24}
\right\}
(A,B,C;p_1,p_2) \nonumber \\
&=& 16\pi^2 \mu^{2\epsilon}
\int \frac{d^n k}{i(2\pi)^n}\frac{k^\mu k^\nu}
{[k^2-m_A^2+i\epsilon][(k+p_1)^2-m_B^2+i\epsilon][(k+p_1+p_2)^2-m_C^2+i\epsilon]},
\nonumber 
\\
\end{eqnarray}
where we use dimensional regularization in $4-2\epsilon$ dimensions, 
and $\mu$ is a renormalization scale.

We list some explicit expressions in the following, where $\frac{1}{\Delta}=\frac{1}{\epsilon}-\gamma+\log 4\pi$:
\begin{eqnarray}
A(m^2)&=&
m^2\left(
\frac{1}{\Delta}+1-\log\frac{m^2}{\mu^2}
\right),\\
B_0(A,B;p)&=& \frac{1}{\Delta}-\int_0^1 dx \log
\frac{m_A^2(1-x)+m_B^2x -p^2x(1-x)-i\epsilon}{\mu^2},\\
B_1(A,B;p)&=&
-\frac{1}{2\Delta}+\int_0^1 dx x\log
\frac{m_A^2(1-x)+m_B^2x -p^2x(1-x)-i\epsilon}{\mu^2},\\
B_{21}(A,B;p)&=&
\frac{1}{3\Delta}-\int_0^1 dx x^2\log
\frac{m_A^2(1-x)+m_B^2x -p^2x(1-x)-i\epsilon}{\mu^2},\\
B_{22}(A,B;p)&=&
\frac{1}{4}(m_A^2+m_B^2-\frac{p^2}{3})\left(\frac{1}{\Delta}+1\right)
\nonumber \\
&&\hspace{-4cm}
-\frac{1}{2}\int_0^1dx \left\{
m_A^2(1-x)+m_B^2 x-p^2x(1-x)
\right\}\log
\frac{m_A^2(1-x)+m_B^2x -p^2x(1-x)-i\epsilon}{\mu^2}.
\end{eqnarray}



\begin{thebibliography}{99}

\bibitem{Aad:2012tfa} 
  G.~Aad {\it et al.}  [ATLAS Collaboration],
  Phys.\ Lett.\ B {\bf 716}, 1 (2012).

\bibitem{Chatrchyan:2012ufa} 
  S.~Chatrchyan {\it et al.}  [CMS Collaboration],
  Phys.\ Lett.\ B {\bf 716}, 30 (2012).

%
%
%

\bibitem{Beringer:1900zz} 
  J.~Beringer {\it et al.}  [Particle Data Group Collaboration],
  Phys.\ Rev.\ D {\bf 86}, 010001 (2012) and 2013 partial update for the 2014 edition.

\bibitem{Aoyama:2012wk} 
  T.~Aoyama, M.~Hayakawa, T.~Kinoshita and M.~Nio,
  Phys.\ Rev.\ Lett.\  {\bf 109}, 111808 (2012)
  [arXiv:1205.5370 [hep-ph]].

\bibitem{Czarnecki:2002nt} 
  A.~Czarnecki, W.~J.~Marciano and A.~Vainshtein,
  Phys.\ Rev.\ D {\bf 67}, 073006 (2003)
  [Erratum-ibid.\ D {\bf 73}, 119901 (2006)].

\bibitem{Hagiwara:2011af} 
  K.~Hagiwara, R.~Liao, A.~D.~Martin, D.~Nomura and T.~Teubner,
  J.\ Phys.\ G {\bf 38}, 085003 (2011)
  [arXiv:1105.3149 [hep-ph]].

\bibitem{Teubner:2010ah} 
  T.~Teubner, K.~Hagiwara, R.~Liao, A.~D.~Martin and D.~Nomura,
  Chin.\ Phys.\ C {\bf 34}, 728 (2010)
  [arXiv:1001.5401 [hep-ph]].

\bibitem{Benayoun:2011mm} 
  M.~Benayoun, P.~David, L.~DelBuono and F.~Jegerlehner,
  Eur.\ Phys.\ J.\ C {\bf 72}, 1848 (2012)
  [arXiv:1106.1315 [hep-ph]].

\bibitem{Jegerlehner:2011ti} 
  F.~Jegerlehner and R.~Szafron,
  Eur.\ Phys.\ J.\ C {\bf 71}, 1632 (2011)
  [arXiv:1101.2872 [hep-ph]].

\bibitem{Jegerlehner:2009ry} 
  F.~Jegerlehner and A.~Nyffeler,
  Phys.\ Rept.\  {\bf 477}, 1 (2009)
  [arXiv:0902.3360 [hep-ph]].

\bibitem{Davier:2010nc} 
  M.~Davier, A.~Hoecker, B.~Malaescu and Z.~Zhang,
  Eur.\ Phys.\ J.\ C {\bf 71}, 1515 (2011)
  [Erratum-ibid.\ C {\bf 72}, 1874 (2012)].

\bibitem{Moroi:1995yh} 
  T.~Moroi,
  Phys.\ Rev.\ D {\bf 53}, 6565 (1996)
  [Erratum-ibid.\ D {\bf 56}, 4424 (1997)].

\bibitem{Hambye:2006zn} 
For example, see
  T.~Hambye, K.~Kannike, E.~Ma and M.~Raidal,
  Phys.\ Rev.\ D {\bf 75}, 095003 (2007)
  [hep-ph/0609228];
  S.~Kanemitsu and K.~Tobe,
  Phys.\ Rev.\ D {\bf 86}, 095025 (2012)
  [arXiv:1207.1313 [hep-ph]].

\bibitem{Abbiendi:2003dh}
  G.~Abbiendi {\it et al.}  [OPAL Collaboration],
  Eur.\ Phys.\ J.\ C {\bf 33}, 173 (2004)
  [hep-ex/0309053].

\bibitem{Abdallah:2005ph}
  J.~Abdallah {\it et al.}  [DELPHI Collaboration],
  Eur.\ Phys.\ J.\ C {\bf 45}, 589 (2006)
  [hep-ex/0512012].

\bibitem{Aaltonen:2008ah}
  T.~Aaltonen {\it et al.}  [CDF Collaboration],
  Phys.\ Rev.\ Lett.\  {\bf 102}, 091805 (2009)
  [arXiv:0811.0053 [hep-ex]].

\bibitem{Abazov:2010ti}
  V.~M.~Abazov {\it et al.}  [D0 Collaboration],
  Phys.\ Lett.\ B {\bf 695}, 88 (2011).

\bibitem{Aad:2012hf}
  G.~Aad {\it et al.}  [ATLAS Collaboration],
  JHEP {\bf 1211}, 138 (2012)
  [arXiv:1209.2535 [hep-ex]].

\bibitem{Chatrchyan:2012oaa}
  S.~Chatrchyan {\it et al.}  [CMS Collaboration],
  Phys.\ Lett.\ B {\bf 720}, 63 (2013)
  [arXiv:1212.6175 [hep-ex]].

\bibitem{Fayet:2007ua} 
  P.~Fayet,
  Phys.\ Rev.\ D {\bf 75}, 115017 (2007)
  [hep-ph/0702176 [HEP-PH]].

\bibitem{Pospelov:2008zw} 
  M.~Pospelov,
  Phys.\ Rev.\ D {\bf 80}, 095002 (2009)
  [arXiv:0811.1030 [hep-ph]].

\bibitem{Endo:2012hp} 
  M.~Endo, K.~Hamaguchi and G.~Mishima,
  Phys.\ Rev.\ D {\bf 86}, 095029 (2012)
  [arXiv:1209.2558 [hep-ph]].

\bibitem{Davoudiasl:2012ig} 
  H.~Davoudiasl, H.~-S.~Lee and W.~J.~Marciano,
  Phys.\ Rev.\ D {\bf 86}, 095009 (2012)
  [arXiv:1208.2973 [hep-ph]].

\bibitem{He:1991qd} 
  X.~-G.~He, G.~C.~Joshi, H.~Lew and R.~R.~Volkas,
  Phys.\ Rev.\ D {\bf 44}, 2118 (1991).

\bibitem{Baek:2001kca} 
  S.~Baek, N.~G.~Deshpande, X.~G.~He and P.~Ko,
  Phys.\ Rev.\ D {\bf 64}, 055006 (2001)
  [hep-ph/0104141].

\bibitem{Ma:2001md} 
  E.~Ma, D.~P.~Roy and S.~Roy,
  Phys.\ Lett.\ B {\bf 525}, 101 (2002)
  [hep-ph/0110146].

\bibitem{Salvioni:2009jp} 
  E.~Salvioni, A.~Strumia, G.~Villadoro and F.~Zwirner,
  JHEP {\bf 1003}, 010 (2010).

\bibitem{Heeck:2011wj} 
  J.~Heeck and W.~Rodejohann,
  Phys.\ Rev.\ D {\bf 84}, 075007 (2011)
  [arXiv:1107.5238 [hep-ph]].

\bibitem{Carone:1994aa} 
  C.~D.~Carone and H.~Murayama,
  Phys.\ Rev.\ Lett.\  {\bf 74}, 3122 (1995)
  [hep-ph/9411256];
  Phys.\ Rev.\  {\bf D52}, 484 (1995)
  [hep-ph/9504393].

\bibitem{Cho:2011rk} 
  G.~-C.~Cho {\it et al.},
  JHEP {\bf 1111}, 068 (2011)
  [arXiv:1104.1769].
\bibitem{Cho:1999km} 
  G.~-C.~Cho and K.~Hagiwara,
  Nucl.\ Phys.\ B {\bf 574}, 623 (2000)
  [hep-ph/9912260].
\bibitem{Hagiwara:1994pw} 
  K.~Hagiwara {\it et al.}, 
  Z.\ Phys.\ C {\bf 64}, 559 (1994)
  [Erratum-ibid.\ C {\bf 68}, 352 (1995)]
  [hep-ph/9409380].
\bibitem{ALEPH:2005ab} 
  S.~Schael {\it et al.}  [ALEPH and DELPHI and L3 and OPAL and SLD and LEP Electroweak Working Group and SLD Electroweak Group and SLD Heavy Flavour Group Collaborations],
  Phys.\ Rept.\  {\bf 427}, 257 (2006)
  [hep-ex/0509008].


\bibitem{Belyaev:2012qa} 
  A.~Belyaev, N.~D.~Christensen and A.~Pukhov,
  Comput.\ Phys.\ Commun.\  {\bf 184}, 1729 (2013)
  [arXiv:1207.6082 [hep-ph]].
\bibitem{Sjostrand:2006za} 
  T.~Sjostrand, S.~Mrenna and P.~Z.~Skands,
  JHEP {\bf 0605}, 026 (2006)
  [hep-ph/0603175].
\bibitem{Ovyn:2009tx} 
  S.~Ovyn, X.~Rouby and V.~Lemaitre,
  arXiv:0903.2225 [hep-ph].
 
\bibitem{CMS:2012bw} 
  S.~Chatrchyan {\it et al.}  [CMS Collaboration],
  JHEP {\bf 1212}, 034 (2012)
  [arXiv:1210.3844 [hep-ex]].

\bibitem{ATLAS_Z4l}
ATLAS Collaboration,
``ATLAS measurements of the 7 and 8 TeV cross sections for $Z\rightarrow 4l$ in pp collisions'',
ATLAS-CONF-2013-055 (May 27, 2013).
 
\bibitem{LHC14trigger}
K. Nakamura, talk at ``Summer camp on ILC accelerator 
and physics/detectors 2013'', Toyama, Japan, July 2013.
 
\bibitem{ATLAS:2013nma} 
  ATLAS Collaboration,
  ATLAS-CONF-2013-013.


\bibitem{POWHEG} 
  P.~Nason,
  JHEP {\bf 0411}, 040 (2004);
  S.~Frixione, P.~Nason and C.~Oleari,
  JHEP {\bf 0711}, 070 (2007);
  S.~Alioli, P.~Nason, C.~Oleari and E.~Re,
  JHEP {\bf 1006}, 043 (2010);
  T.~Melia, P.~Nason, R.~Rontsch and G.~Zanderighi,
  JHEP {\bf 1111}, 078 (2011).



\bibitem{seesaw}
 T. ~Yanagida, in ``Proceedings of the Workshop on Unified Theory and Baryon Number of the Universe,'' eds;  O. Sawada and A. Sugamoto (KEK, 1979) p.95;
 M. Gell- Mann, P. Ramond and R. Slansky, in ``Supergravity,''
 eds.; P. van Niewwenhuizen and D. Freedman (North Holland, Amsterdam,
 1979). See also 
  P.~Minkowski,
  Phys.\ Lett.\  {\bf B67}, 421 (1977).

\bibitem{Kuzmin:1985mm} 
  V.~A.~Kuzmin, V.~A.~Rubakov and M.~E.~Shaposhnikov,
  Phys.\ Lett.\ B {\bf 155}, 36 (1985).


\bibitem{'tHooft:1978xw} 
  G.~'t Hooft and M.~J.~G.~Veltman,
  Nucl.\ Phys.\ B {\bf 153}, 365 (1979);
  G.~Passarino and M.~J.~G.~Veltman,
  Nucl.\ Phys.\ B {\bf 160}, 151 (1979).


\end{thebibliography}
\end{document}